\documentclass[prx, twocolumn, superscriptaddress]{revtex4}
\usepackage{graphicx,color}
\usepackage{amssymb}
\usepackage{amsmath}
\usepackage{bm}
\usepackage{hyperref}
\usepackage{tabularx}
\usepackage{multirow}
\usepackage{braket}
\usepackage{xcolor}
\usepackage{gensymb}

\begin{document}

\title{Magnetic Impurity as a Local Probe of the $U\left(1\right)$ Quantum Spin Liquid with Spinon Fermi Surface}
\author{Wen-Yu He} \thanks{hewy@shanghaitech.edu.cn}
\affiliation{School of Physical Science and Technology, ShanghaiTech University, Shanghai 201210, China}
\affiliation{Department of Physics, Massachusetts Institute of Technology, Cambridge, Massachusetts 02139, USA}
\author{Patrick A. Lee} \thanks{palee@mit.edu}
\affiliation{Department of Physics, Massachusetts Institute of Technology, Cambridge, Massachusetts 02139, USA}

\date{\today}
\pacs{}

\begin{abstract}
We solve the problem of a magnetic impurity coupled to a $U\left(1\right)$ quantum spin liquid with spinon Fermi surface and compute the impurity spectral function. Using the slave rotor mean field approach combined with gauge field fluctuations, we find that a peak located at the top of the lower Hubbard band and one at the bottom of the upper Hubbard band can emerge in the impurity spectral function. The peaks at the Hubbard band edges arise from the gauge field fluctuations induced spinon chargon binding inside the spinon Kondo screening cloud of the magnetic impurity. For a magnetic impurity embedded in a Mott insulator, our findings suggest that the emergence of a pair of peaks at the Hubbard band edges in the impurity density of states spectra  provides strong evidence that the host Mott insulator is a $U\left(1\right)$ quantum spin liquid with spinon Fermi surface.
\end{abstract}

\maketitle

\section{Introduction}
The Anderson impurity model describes how a local atomic impurity interacts with electrons in its host system~\cite{Anderson1}. The most well studied case is a magnetic impurity embedded in a nonmagnetic metal, where the Coulomb repulsion at the impurity suppresses the charge fluctuations but the local magnetic moment can freely couple with spins of nearby conducting electrons in the metal~\cite{Hewson}. The Anderson model of a magnetic impurity can be reduced to the Kondo model~\cite{Hewson, JunKondo} that focuses on the spin interaction between the impurity and the nearby electrons.  Below the Kondo temperature $T_{\textrm{K}}$, the conducting electron spins form the Kondo cloud to screen the local magnetic moment~\cite{Affleck, Zawadowski, Scalapino, Affleck2, Yamamoto}, and a correlated electronic state develops at the impurity. The correlated state is characterized by a narrow peak, known as the Kondo resonance~\cite{Hewson, Zawadowski}, which always shows up at the Fermi energy in the impurity electronic spectral function. Therefore, the Kondo resonance has become the fingerprint for the correlation phenomenon induced by a magnetic impurity on a metallic host~\cite{Delley, Wingreen}.

The thorough study of a magnetic impurity on a metallic host has stimulated further interest to generalize the noninteracting metallic host to a quantum system with strong electron electron interaction, such as quantum spin liquid (QSL)~\cite{Florens1, Sachedev, Kim, Ribeiro, Zorko, Lado, Baigeng, Shichao}. As the magnetic impurity couples with the electronic states locally in the QSL host, the correlated many-body ground state in the QSL is expected to affect the quantum state at the impurity. In this way the magnetic impurity acts as a local probe of the QSL. Among QSLs with various gauge groups~\cite{Lee1, Balents, Ng}, the $U\left(1\right)$ QSL with spinon Fermi surface (SFS) is particularly interesting, and recently TaS$_2$ and monolayer TaSe$_2$ have been studied as the candidates for the $U\left(1\right)$ QSL with SFS~\cite{Law, Lee2, Gang2, Arcon, Yayu, Butler1, Butler2, Yeom, Crommie1, Crommie2}. Furthermore, it has been shown to be experimentally possible to deposit magnetic impurities on the  monolayer TaSe$_2$ surface and perform local spectroscopy measurements on the magnetic impurity~\cite{Crommie4}. Since scanning tunneling spectroscopy (STS) has been proven to be a direct experimental way~\cite{Crommie3} to study the physics of a magnetic impurity on an electronic Fermi liquid metal, the analogous problem of the STS spectrum when a magnetic impurity is coupled to a $U\left(1\right)$ QSL with SFS, becomes a key problem that needs to be theoretically solved. While the possibility of Kondo screening by a spinon Fermi surface has previously been addressed~\cite{Ribeiro}, the tunneling density of states have not been calculated. This is the main goal of this paper.

The $U\left(1\right)$ QSL with SFS is a Mott insulator with no magnetic order down to zero temperature~\cite{Lee1, Balents, Ng}. In the $U\left(1\right)$ QSL with SFS, the electrons go through the spin-charge separation and are decomposed into spinons and chargons, with an emergent $U\left(1\right)$ gauge field to couple the both~\cite{Lee3}. The spinons are charge neutral spin-1/2 excitations living on a Fermi surface, while the chargons are gapped spinless bosons that carry electric charges. For a magnetic impurity embedded on a $U\left(1\right)$ QSL with SFS, since the spinons in the QSL are itinerant, the magnetic impurity can have spin exchange freely with the itinerant spinons nearby and the Kondo model that involves the spin degree coupling has been applied to study the system~\cite{Ribeiro}.

\begin{figure*}
\centering
\includegraphics[width=6.8in]{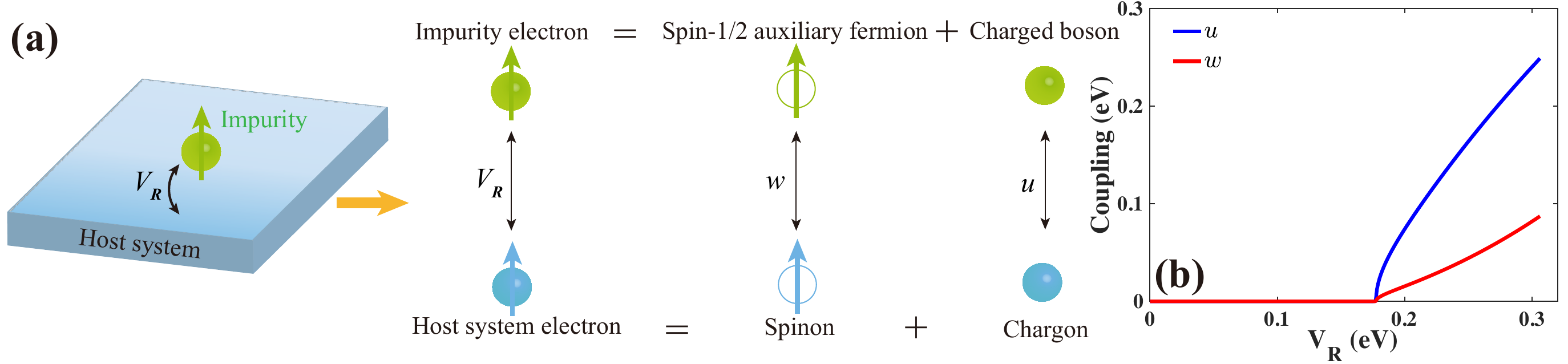}
\caption{(a) A magnetic impurity embedded in a host system and the slave rotor mean field description for the system. In the slave rotor representation, the electronic state at the magnetic impurity is decomposed into a SAF and a CB, while the electrons in the host system are decomposed into the spinons and chargons. The SAF has the mean field coupling $w$ with the spinons, and the CB takes the mean field coupling $u$ with the chargons. The mean field couplings $w$ and $u$ are developed from the orginal electronic coupling $V_{\bm{R}}$. (b) The mean field coupling parameters $w$ and $u$ as a function of the electronic coupling $V_{\bm{R}}$ for a magnetic impurity on a $U\left(1\right)$ QSL with SFS. The QSL is of trianglar lattice. As $V_{\bm{R}}$ increases, both $w$ and $u$ take significant nonzero values.}\label{figure1}
\end{figure*}

In the Kondo model description for the magnetic impurity on a $U\left(1\right)$ QSL with SFS, the itinerant spinons in the QSL take the role of the spin-1/2 conducting electrons in the metal, so it was predicted and further supported by experimental evidence that a spinon cloud can form around the magnetic impurity to screen the local magnetic moment~\cite{Ribeiro, Zorko}, similar to the conventional Kondo phenomenon in a metallic host. However, unlike the conventional Kondo problem where the impurity magnetic moment couples to the spin of charged electrons in the metallic host, the Kondo model adopted in the $U\left(1\right)$ QSL host only takes into account the spinons that carry no charge. Since the local spectroscopy measurements on the magnetic impurity involves the physical electron transfer that contains both the spin and charge degree, the Kondo model used in the QSL host~\cite{Ribeiro} becomes insufficient to generate the impurity electronic spectral function. 

In this work, we obtain the impurity electronic spectral function of a magnetic impurity embedded in a $U\left(1\right)$ QSL with SFS by solving the Anderson impurity model, which considers both the spin and charge degree of electrons. The slave rotor mean field theory~\cite{Florens2, Florens3} is applied to deal with the Anderson impurity model. At the magnetic impurity, the electron is decomposed into a spin-1/2 auxiliary fermion (SAF) and a charged boson (CB), which couples with the spinons and chargons in the QSL respectively. The coupling of the SAF and the spinons corresponds to the Kondo model that contains only the spin degree, generating the resonance at the spinon Fermi energy in the SAF spectral function. The coupling of the CB and the chargons fills in the missing charge degree of the Kondo model. At the mean field level with no gauge field fluctuations in the QSL, the impurity electronic spectral function obtained from the slave rotor treatment involves the QSL chargon density of states (DOS) in addition to the original atomic energy levels, so the magnetic impurity on a $U\left(1\right)$ QSL with SFS locally provides information on the QSL chargon DOS. 

In a $U\left(1\right)$ QSL, the spinons and chargons both couple to an emergent $U\left(1\right)$ gauge field, so the quantum fluctuations of the $U\left(1\right)$ gauge field affect the low energy excitations in the QSL. In a two-dimensional $U\left(1\right)$ QSL with gapped spinon spectrum, it is known from Polyakov that the low energy effective theory of the QSL is a pure compact $U\left(1\right)$ gauge theory which is always confining. As a result, an arbitrary coupling to the $U\left(1\right)$ gauge field gives rise to confinement of spinons~\cite{Polyakov}. However, in a $U\left(1\right)$ QSL with gapless spinon spectrum, since the gapless matter field damps the gauge field, the situation becomes different. In a $U\left(1\right)$ QSL with a Dirac spinon band dispersion, it has been found that when the physical $SU\left(2\right)$ spin is generalized to $SU\left(N\right)$, deconfinement of spinons is possible for a sufficiently large $N$~\cite{Xiaogang}. In our case of a two-dimensional $U\left(1\right)$ QSL with SFS, due to the many gapless spinon excitations from the SFS, a noncompact $U\left(1\right)$ gauge theory applies to describe the low energy excitations, and it becomes possible to have deconfinement of spinons~\cite{SungSikLee}. In the rest of the paper, deconfinement of spinons is a prerequisite for the treatment of gauge field fluctuations that we apply to the two-dimensional $U\left(1\right)$ QSL with SFS.

In a two-dimensional $U\left(1\right)$ QSL with SFS, as the spinons and chargons both couple to an emergent $U\left(1\right)$ gauge field, the gauge field fluctuations can generate a binding interaction between a spinon and a chargon~\cite{Lee4, Lee5}. In the spinon Kondo cloud around the magnetic impurity, the spinon chargon binding interaction is found to induce the spectral weight to transfer from the bulk Hubbard bands to the Hubbard band edges, so a band edge resonance peak arises at the top of the lower Hubbard band (LHB) and one arises at the bottom of the upper Hubbard band (UHB) in the impurity spectral function. When the spinon chargon binding interaction is sufficiently large, the pair of band edge resonance peaks can further move inside the Mott gap and become in-gap bound state peaks. Due to the spinon Kondo cloud of the magnetic impurity and the spinon chargon binding interaction inside, the magnetic impurity embedded on a $U\left(1\right)$ QSL with SFS behaves like an impurity of acceptor and donor type simultaneously and induces a pair of peaks at the Hubbard band edge energies. As a result, in the spectroscopy of the magnetic impurity, the emergence of a peak at both the UHB bottom and LHB top gives the strong evidence that the host Mott insulator is a $U\left(1\right)$ QSL with SFS.

The rest of the paper is organized as follows: In section \ref{II}, the slave rotor mean field theory to solve the Anderson impurity model is introduced. In section \ref{III}, a magnetic impurity weakly coupled to a $U\left(1\right)$ QSL with SFS is analyzed in both the SAF spinon coupling channel and the CB chargon coupling channel. The impurity bare electronic spectral function, which does not consider the gauge field fluctuations, is analytically derived. The resulting impurity bare electronic spectral function is shown to measure the chargon DOS in the QSL host. In section \ref{VI}, full numerical self-consistent mean field calculation for a magnetic impurity embedded in a $U\left(1\right)$ QSL with SFS of triangular lattice is carried out. The impurity bare electronic spectral function obtained numerically confirms that a magnetic impurity on a $U\left(1\right)$ QSL with SFS probes the chargon DOS in the QSL, under the condition that the QSL has negligible gauge field fluctuations. In section \ref{V}, the gauge field flutuations in the QSL are taken into account and the effect is to bring about the spinon chargon binding interaction in the QSL. Around the magnetic impurity, the binding of a spinon state in the spinon Kondo cloud and a chargon makes the magnetic impurity become an impurity of acceptor and donor type simultaneously, giving rise to a peak located at the UHB bottom and one at the LHB top in the impurity spectral function. The pair of peaks are either band edge resonance peaks inside the Hubbard bands or in-gap peaks of the formed bound states, depending on the binding interaction strength. In section \ref{VI}, our findings about the magnetic impurity on a $U\left(1\right)$ QSL with SFS are summarized and the connection to experimental spectroscopy measurements of a magnetic impurity on a Mott insulator is constructed.

\section{The Slave Rotor Mean Field Theory for the Impurity Model}\label{II}
The Hamiltonian for a local atomic impurity with onsite Coulomb repulsion $U$ takes the form
\begin{align}\nonumber
H_{\textrm{d}}=&\sum_\sigma\epsilon_0 d^\dagger_\sigma d_\sigma+\frac{U}{2}\left(\sum_\sigma d^\dagger_\sigma d_\sigma-1\right)^2.
\end{align}
Here $\epsilon_0$ is the impurity onsite energy, and $\sigma=\uparrow/\downarrow$ is the spin index. In the slave rotor formalism~\cite{Florens2, Florens3}, the electronic operator at the impurity is rewritten as $d^\dagger_\sigma=a^\dagger_\sigma e^{i\theta}$, $d_\sigma=a_\sigma e^{-i\theta}$, where $a^{\left(\dagger\right)}_\sigma$ is the annihilation (creation) operator for a SAF and $e^{\pm i\theta}$ is the rotor to represent the charge degree of the impurity electron. In the slave rotor representation, the angular momentum $L=-i\frac{\partial}{\partial\theta}$ is introduced to express the Coulomb repulsion term in $H_d$ as $\frac{U}{2}L^2$, under the constraint $L=\sum_\sigma a^\dagger_\sigma a_\sigma-1$. As the angular momentum $L$ and the rotor phase variable $\theta$ respect the canonical commutation $\left[\theta, L\right]=i$, the action for the impurity can be constructed from $\mathcal{S}=\int_0^\beta\left(-iL\partial_\tau\theta+\sum_\sigma a^\dagger_\sigma a_\sigma+H_d\right)d\tau$. After an integration over $L$, we get the action for the local atomic impurity as~\cite{Florens2}
\begin{align}\nonumber
S_{\textrm{d}}=&\int_0^\beta\left[\frac{1}{U}\left(\partial_\tau+h\right)X^\dagger_d\left(\partial_\tau-h\right)X_d\right.\\
&\left.+\sum_\sigma a^\dagger_\sigma\left(\partial_\tau+\epsilon_0-h\right)a_\sigma+\lambda\left(X^\dagger_dX_d-1\right)+h\right]d\tau,
\end{align}
with $X^{\dagger}_d=e^{-i\theta}$, $X_d=e^{i\theta}$ interpreted as the charged bosonic operator at the impurity, $\lambda$ the Lagrangian multiplier to guarantee $X^\dagger_dX_d=1$, and $h$ the Lagrangian multiplier to implement the constraint $L=\sum_\sigma a^\dagger_\sigma a_\sigma-1$. In the action $S_{\textrm{d}}$, the impurity electron has been decomposed into a SAF and a CB, which can couple with external spin and charge degrees individually.

Embedded on a host system, the atomic impurity couples with the host, yielding the coupling action
\begin{align}
S_{\textrm{c}}=\int_0^\beta\sum_\sigma V_{\bm{R}}\left(f^\dagger_{\sigma, \bm{R}}a_\sigma X^\dagger_d X_{\bm{R}}+a^\dagger_\sigma f_{\sigma, \bm{R}}X^\dagger_{\bm{R}}X_d\right)d\tau, 
\end{align}
where $V_{\bm{R}}$ is the electronic coupling with the nearest neighbor site $\bm{R}$ in the host system. Here the electrons in the host system have been expressed in the slave rotor representation as well: $c^\dagger_{\sigma, \bm{r}}=f^\dagger_{\sigma, \bm{r}}X_{\bm{r}}$, $c_{\sigma, \bm{r}}=f_{\sigma, \bm{r}}X^\dagger_{\bm{r}}$, with $f^{\left(\dagger\right)}_{\sigma, \bm{r}}$, $X^{\left(\dagger\right)}_{\bm{r}}$ to annihilate (create) a spinon and chargon respectively at the site $\bm{r}$. The spinon is a chargeless spin-1/2 fermion and the chargon is a spinless CB, representing the spin degree and the charge degree of an electron in the host respectively. With the imaginary time Green's function $G_{f, \sigma}\left(\tau, \tau', \bm{r}, \bm{r}'\right)$ and $G_X\left(\tau, \tau', \bm{r}, \bm{r}'\right)$ for the spinon and chargon respectively, the action for the host system can be written as
\begin{align}\nonumber
S_{\textrm{h}}=&-\int_{\tau, \tau', \bm{r}, \bm{r}'}\sum_\sigma f^\dagger_{\sigma, \bm{r}}\left(\tau\right)G^{-1}_{f, \sigma}\left(\tau, \tau', \bm{r}, \bm{r}'\right)f_{\bm{r}', \sigma}\left(\tau'\right)\\
&+\int_{\tau, \tau', \bm{r}, \bm{r}'}X^\dagger_{\bm{r}}\left(\tau\right)G^{-1}_X\left(\tau, \tau', \bm{r}, \bm{r}'\right)X_{\bm{r}'}\left(\tau'\right),
\end{align}
with $\int_{\tau, \tau', \bm{r}, \bm{r}'}\equiv\int_0^\beta d\tau\int_0^\beta d\tau'\int d\bm{r}\int d\bm{r}'$. 

Now it is ready to have the total action for an atomic impurity embedded on a host system as $S_{\textrm{tot}}=S_{\textrm{d}}+S_{\textrm{c}}+S_{\textrm{h}}$. In the total action $S_{\textrm{tot}}$, all terms are quadratic except the ones in $S_{\textrm{c}}$, so the mean field parameters $w$ for the SAF spinon coupling, $u$ for the CB chargon coupling are introduced through the Hubbard-Stratonovich transformation to decouple the quartic terms in $S_{\textrm{c}}$. Eventually, the action for the whole system takes the form
\begin{widetext}
\begin{align}\nonumber\label{Act_tot}
S_0=&\int_0^\beta\left[\sum_\sigma a^\dagger_\sigma\left(\partial_\tau+\epsilon_0-h\right)a_\sigma+\frac{\partial_\tau X_d^\dagger\partial_\tau X_d}{U}+\frac{h}{U}\left(X^\dagger_d\partial_\tau X_d-X_d\partial_\tau X^\dagger_d\right)+\lambda\left(X^\dagger_dX_d-1\right)+h-\frac{h^2}{U}\right]d\tau\\\nonumber
&-\int_0^\beta d\tau\int_0^\beta d\tau'\int d\bm{r}\int d\bm{r}'\sum_\sigma\left[f^\dagger_{\sigma, \bm{r}}\left(\tau\right)G^{-1}_{f, \sigma}\left(\tau, \tau', \bm{r}, \bm{r}'\right)f_{\bm{r}', \sigma}-X^\dagger_{\bm{r}}\left(\tau\right)G^{-1}_X\left(\tau, \tau', \bm{r}, \bm{r}'\right)X_{\bm{r}'}\left(\tau\right)\right]\\
&+\int_0^\beta\left[\frac{2uw}{V_{\bm{R}}}+w\left(f^\dagger_{\sigma, \bm{R}}a_\sigma+a^\dagger_\sigma f_{\sigma, \bm{R}}\right)-u\left(X^\dagger_dX_{\bm{R}}+X^\dagger_{\bm{R}}X_d\right)\right]d\tau.
\end{align}
\end{widetext}
In this slave rotor treatment, the coupling of the impurity electron and the electronic states in the host has been decomposed into two individual channels: the SAF has the coupling $w$ with the spinons, and the CB takes the coupling $u$ with the chargons. The mean field coupling parameters $w$ and $u$ are controlled by the electronic coupling $V_{\bm{R}}$. For an atomic impurity embedded in a host system, the decomposition of electronic coupling into the SAF spinon coupling channel and the CB chargon coupling channel is schematically shown in Fig. \ref{figure1} (a). With given spinon and chargon Green's function in the host system, the action in Eq. \ref{Act_tot} is generally applicable to a local atomic impurity embedded on a host, no matter the host system is a metal or a QSL.

In the mean field theory, the coupling fields $w$, $u$ and the Lagrangian multipliers $\lambda$, $h$ take the value at the saddle point where the free energy for the whole system gets  minimized. After the Fourier transformation to the Matsubara frequency space, the self-consistent equations that determine the value of $w$, $u$, $\lambda$, and $h$ are
\begin{align}\label{SF_Eq1}
u=&-\frac{2wV_{\bm{R}}}{\beta}\sum_{\omega_n}G_{a, \sigma}\left(i\omega_n\right)G_{f, \sigma}\left(i\omega_n, \bm{R}, \bm{R}\right),\\\label{SF_Eq2}
w=&\frac{uV_{\bm{R}}}{\beta}\sum_{\nu_n}G_{X_d}\left(i\nu_n\right)G_X\left(i\nu_n, \bm{R}, \bm{R}\right),\\\label{SF_Eq3}
1=&\frac{1}{\beta}\sum_{\nu_n}G_{X_d}\left(i\nu_n\right)e^{i\nu_n0^+},\\\nonumber\label{SF_Eq4}
\frac{1}{2}-\frac{h}{U}=&\frac{1}{2U\beta}\sum_{\nu_n}i\nu_nG_{X_d}\left(i\nu_n\right)\left[e^{i\nu_n0^+}+e^{-i\nu_n0^+}\right]\\
&+\frac{1}{\beta}\sum_{\omega_n}G_{a, \sigma}\left(i\omega_n\right)e^{i\omega_n0^+},
\end{align}
with the SAF Matsubara Green's function
\begin{align}\label{Green_auxiliary}
G_{a, \sigma}\left(i\omega_n\right)=\frac{1}{i\omega_n-\epsilon_0+h-w^2G_{f, \sigma}\left(i\omega_n, \bm{R}, \bm{R}\right)},
\end{align}
and the CB Matsubara Green's function
\begin{align}\label{Green_boson}
G_{X_d}\left(i\nu_n\right)=\frac{1}{\frac{\nu_n^2}{U}-\frac{2i\nu_nh}{U}+\lambda-u^2G_X\left(i\nu_n, \bm{R}, \bm{R}\right)}.
\end{align}
Here $G_{f, \sigma}\left(i\omega_n, \bm{R}, \bm{R}\right)$ and $G_{X}\left(i\nu_n, \bm{R}, \bm{R}\right)$ are the Matsubara Green's functions for the spinon and chargon at $\bm{R}$ respectively, $\omega_n=\frac{\left(2n+1\right)\pi}{\beta}$ and $\nu_n=\frac{2n\pi}{\beta}$ denote the fermionic and bosonic Matsubara frequencies respectively, and $k_{\textrm{b}}$ in $\beta^{-1}=k_{\textrm{b}}T$ is the Boltzman constant. The first two equations in Eq. \ref{SF_Eq1} and Eq. \ref{SF_Eq2} show how the mean field couplings $w$ and $u$ arise from the electronic coupling $V_{\bm{R}}$, while the last two equations in Eq. \ref{SF_Eq3} and Eq. \ref{SF_Eq4} determine the chemical potential for the SAF and the CB at the impurity.

Importantly, for an atomic impurity with given onsite energy $\epsilon_0$, Coulomb repulsion $U$, electronic coupling $V_{\bm{R}}$ and the Green's functions $G_{f, \sigma}\left(i\omega_n, \bm{R}, \bm{R}\right)$, $G_{X}\left(i\nu_n, \bm{R}, \bm{R}\right)$ for the host system, all the mean field parameters $w$, $u$, $\lambda$ and $h$ can be self-consistently solved from Eq. \ref{SF_Eq1} - \ref{SF_Eq4}. For a magnetic impurity embedded in a $U\left(1\right)$ QSL with SFS that we concern, the mean field equations will be numerically solved to yield the SAF spinon coupling $w$ and the CB chargon coupling $u$ as a function of the electronic coupling $V_{\bm{R}}$, which is shown in Fig. \ref{figure1} (b). At the mean field level where all gauge field fluctuations are neglected, the impurity bare electronic Matsubara Green's function $G^0_{d, \sigma}\left(i\omega_n\right)$ is constructed from the convolution
\begin{align}\label{Convolution}
G^0_{d, \sigma}\left(i\omega_n\right)=\frac{1}{\beta}\sum_{\nu_n}G_{a, \sigma}\left(i\omega_n+i\nu_n\right)G_{X_d}\left(i\nu_n\right),
\end{align}
and the impurity bare electronic spectral function takes the form $\rho^0_d\left(\omega\right)=-\frac{1}{\pi}\textrm{Im}G^0_{d, \sigma}\left(i\omega_n\rightarrow\omega+i0^+\right)$. We refer to the Green's function and spectral function as bare because the spinon-chargon binding has not yet been taken into account. Therefore the bare spectral function $\rho^0_{d, \sigma}\left(\omega\right)$ gives the bare electronic DOS of a magnetic impurity based on a $U\left(1\right)$ QSL with SFS, which is the result in the absence of gauge field fluctuations.

\begin{figure*}
\centering
\includegraphics[width=6.8in]{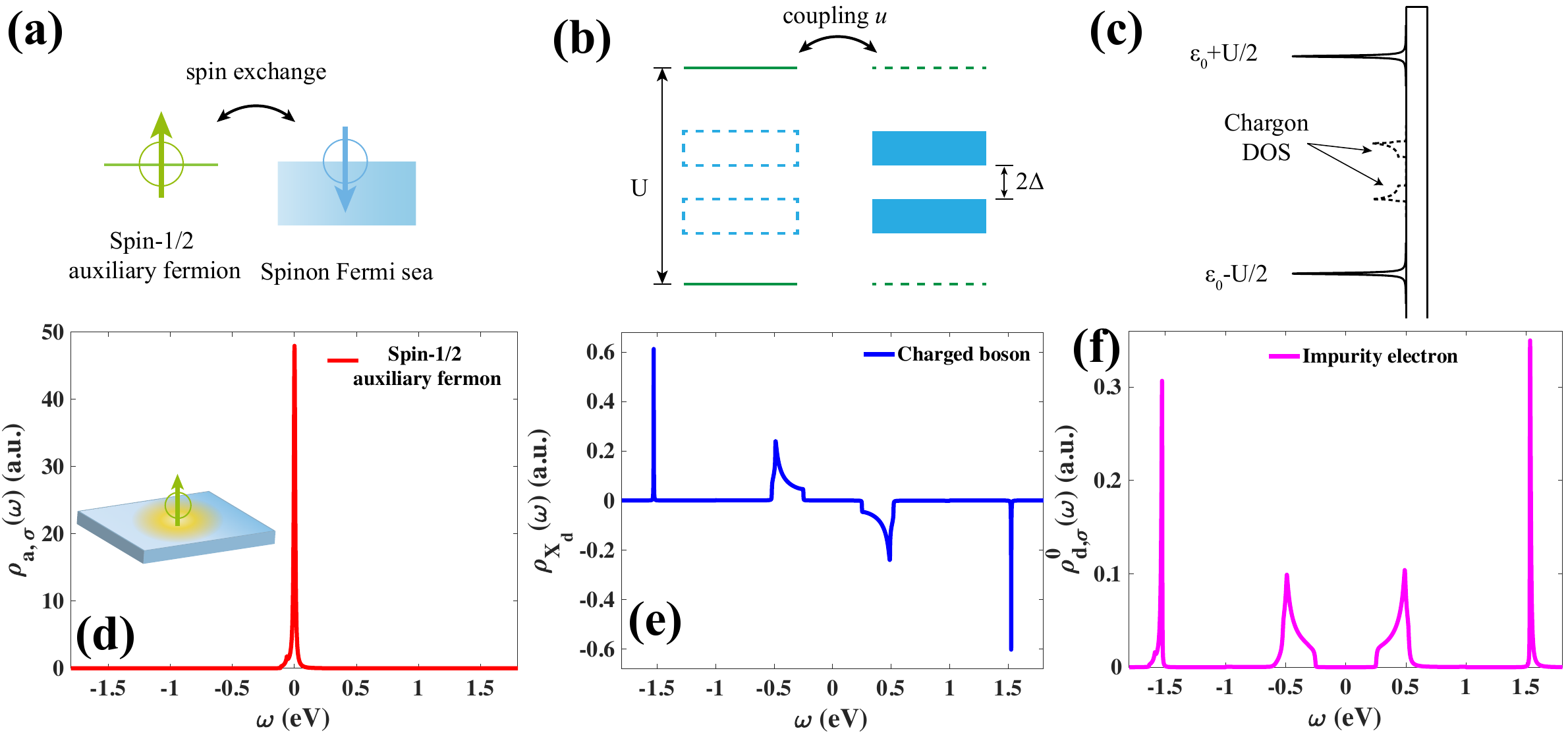}
\caption{(a) The spin exchange between the SAF at the impurity and the itinerant spinons in the $U\left(1\right)$ QSL with SFS. The impurity SAF lies at the Fermi level and can freely exchange spin with the QSL spinons in the Fermi sea. (b) The energy diagram of the CB at the impurity and the chargons in the QSL. In the absence of the mean field coupling $u$, the CB has two energy levels denoted by the solid green lines, which are separated by the Coulomb repulsion $U$. The chargon energy bands are denoted by the solid blue rectangles, which are separated by the chargon gap $2\Delta$. With finite mean field coupling $u$, the hybridization between the CB and the chargons mix the states, so the CB has a fraction of states at the energies of chargon bands, denoted by the blue dashed rectangles. Meanwhile, the hybridized chargons have states at the energy levels of the CB, denoted by the green dashed lines. (c) The impurity electronic DOS diagram. The magnetic impurity has the original atomic energy levels at $\omega=\epsilon_0\pm\frac{U}{2}$, which give the peaks in the solid spectrum. With finite coupling to a $U\left(1\right)$ QSL with SFS, there emerge new states at the Hubbard band energies denoted by the dashed spectrum. (d) The resonance peak at the Fermi energy in the SAF spectral function $\rho_{a, \sigma}\left(\omega\right)$. The inset shows the local spin at the impurity is screened by the spinon Kondo cloud, which is in gold color. (e) The CB spectral function $\rho_{X_d}\left(\omega\right)$. (f) The impurity bare electronic spectral function $\rho^0_{d, \sigma}\left(\omega\right)$, before spinon-chargon binding is taken into account. The spectral functions in (d), (e), and (f) are numerically calculated for a magnetic impurity on a QSL of triangular lattice.}\label{figure2}
\end{figure*}

\section{The Magnetic Impurity Weakly Coupled to the QSL}\label{III}
The full analytical solution to Eq. \ref{SF_Eq1} - \ref{SF_Eq4} in the presence of nonzero electronic coupling $V_{\bm{R}}$ is difficult to get, but at weak coupling $V_{\bm{R}}$ we can make approximations to first get the impurity bare electronic spectral function. Since the weak coupling to the host system has negligible effect on the occupation of the atomic impurity, the Lagrangian multipliers $h$ and $\lambda$, which determine the chemical potential for the SAF and the CB at the impurity, can be approximated by the value in the isolated magnetic impurity case. 

By taking $u=0$ and $w=0$ in Eq. \ref{SF_Eq3} and Eq. \ref{SF_Eq4}, we obtain $h=\epsilon_0$ and $\lambda=\frac{U}{4}-\frac{h^2}{U}$ under the condition $|\epsilon_0|<\frac{1}{2}U$~\cite{Florens2}, which corresponds to an isolated single occupied atomic impurity that carries a local magnetic moment. The Matsubara Green's function for the SAF and the CB can accordingly be obtained from Eq. \ref{Green_auxiliary} and Eq. \ref{Green_boson} respectively. In such an isolated magnetic impurity, the SAF has its energy level right at the Fermi energy, indicating that the local spin at the impurity can have free exchange with external spins, as is schematically shown in Fig. \ref{figure2} (a). On the other hand, the CB at the impurity has two energy levels gapped by the Coulomb repulsion $U$, denoted by the solid green lines in Fig. \ref{figure2} (b). The CB and the SAF together compose the impurity electron, and the electronic spectrum has the atomic energy levels at $\omega=\epsilon_0\pm\frac{U}{2}$ as is shown schematically by the solid spectrum in Fig. \ref{figure2} (c). The gapped electronic spectrum at the magnetic impurity indicates that extra charging energy is needed to remove or add an electron into the impurity, so the charge fluctuations are suppressed by the gap. However, since the isolated magnetic impurity has the same energy for the spin up and spin down electronic states, the impurity spin can fluctuate freely.

With the approximated Lagrangian multipliers $h=\epsilon_0$ and $\lambda=\frac{U}{4}-\frac{h^2}{U}$ taken from the isolated magnetic impurity model, the coupling $u$ in the CB chargon channel and the coupling $w$ in the spin-1/2 auxliary fermion spinon channel can be determined by Eq. \ref{SF_Eq1} and Eq. \ref{SF_Eq2} respectively. For a $U\left(1\right)$ QSL with SFS as the base for the magnetic impurity, the spinon Green's function takes the form $G_{f, \sigma}\left(i\omega_n, \bm{R}, \bm{R}\right)=\sum_{\bm{k}}\frac{1}{i\omega_n-\xi_{\bm{k}}}$ with the spinon band dispersion $\xi_{\bm{k}}\in\left[-\Lambda_f, \Lambda_f\right]$, and the chargon Green's function is $G_X\left(i\nu_n, \bm{R}, \bm{R}\right)=\sum_{\bm{k}}\left(\frac{1}{i\nu_n+\epsilon_{\bm{k}}}-\frac{1}{i\nu_n-\epsilon_{\bm{k}}}\right)$, given the chargon band dispersion $\epsilon_{\bm{k}}\in\left[\Delta, \Lambda_X\right]$ (see Appendix \ref{App-A} for more details about the Green's function). By assuming the constant DOS for both the spinon and chargon bands for simplicity, the equations in Eq. \ref{SF_Eq1} and Eq. \ref{SF_Eq2} which contain infinite summation series are further approximated by two algebra equations
\begin{align}
u\approx&-\frac{wV_{\bm{R}}}{2\Lambda_f}\ln\frac{\Gamma^2_{\textrm{eff}}}{\Lambda^2_f+\Gamma^2_{\textrm{eff}}},\quad
w\approx\frac{4uV_{\bm{R}}U}{U^2-4\epsilon^2_0},
\end{align}
with the effective broadening function defined as $\Gamma_{\textrm{eff}}=\frac{\pi w^2}{2\Lambda_f}$. The detailed derivation about the two algebra equations are present in Appendix \ref{App-B}. The electronic coupling $V_{\bm{R}}$ thereby determines the mean field couplings $w$ and $u$ in the SAF spinon channel and the CB chargon channel respectively.

In the coupling channel of the SAF and spinon, now the finite mean field coupling $w$ allows auxiliary fermion at the impurity to have spin exchange with the itinerant spinons in the $U\left(1\right)$ QSL, as is shown in Fig. \ref{figure2} (a). Near the impurity, itinerant spinons at the Fermi level get spin flip scattered by the local magnetic moment, and the local spin at the impurity is also coherently switched in the scattering process. This magnetic scattering yields resonance at the impurity. In the presence of finite coupling to the itinerant spinons, the Matsubara Green's function for the SAF at the impurity is modified to be $G_{a, \sigma}\left(i\omega_n\right)=\frac{1}{i\omega_n-i\Gamma_{\textrm{eff}}}$, so the SAF spectral function $\rho_{a, \sigma}\left(\omega\right)=\frac{1}{\pi}\frac{\Gamma_{\textrm{eff}}}{\omega^2+\Gamma^2_{\textrm{eff}}}$ always shows a resonance at the Fermi energy. This is the Kondo like resonance in the spinon channel. The width of the resonance is solved to be $\Gamma_{\textrm{eff}}=\Lambda_f\exp\left(-\frac{U\Lambda_f}{4V^2_{\bm{R}}}\right)$, which denotes the Kondo temperature energy scale $k_{\textrm{b}}T_{\textrm{K}}$ same as that in a metallic host~\cite{Hewson}. Below the Kondo temperature $T_{\textrm{K}}$, the nearby spinons of opposite spin are strongly correlated with the local SAF at the impurity, forming the Kondo cloud of size $\xi_{\textrm{K}}$~\cite{Affleck} to screen the local magnetic moment, as is shown in Fig. \ref{figure2} (d) inset. For a magnetic impuity based on a $U\left(1\right)$ QSL with SFS, the spinon Kondo resonance and the spinon Kondo screening cloud are in analogy to the conventional Kondo effect of a magnetic impurity in a metal, as in both cases the spin exchange enables the spin fluctuations at the impurity.

The CB chargon coupling channel of the magnetic impurity based on a $U\left(1\right)$ QSL, on the other hand, makes the difference from that in conventional Kondo effect. In the QSL, the chargons are gapped, and the mean field coupling $u$ enables the hybridization between the chargons and the CB at the impurity. Due to the hybridization, the charged bosonic state at the impurity gets mixed with the chargon states in the QSL, so a fraction of chargon states are transferred to the CB at the impurity, as is schematically shown in Fig. \ref{figure2} (b) by the blue dashed rectangles. The hybridization with the chargons in the QSL modifies the CB Matsubara Green's function to be $G_{X_d}\left(i\nu_n\right)\approx\frac{1}{i\nu_n+\epsilon_0+\frac{U}{2}}-\frac{1}{i\nu_n+\epsilon_0-\frac{U}{2}}+ZG_X\left(i\nu_n, \bm{R}, \bm{R}\right)$ with $Z=\frac{16u^2U^2}{\left(U^2-4\epsilon_0^2\right)^2}$ the spectral weight (see Appendix \ref{App-C} for the derivation about the spectral weight), so the CB spectral function $\rho_{X_d}\left(\omega\right)=-\frac{1}{\pi}\textrm{Im}G_{X_d}\left(i\nu_n\rightarrow\omega+i0^+\right)$ now not only shows the peaks from the original two energy levels, but also includes a fraction of chargon states at the energy of QSL chargon bands. Differing from the conventional Kondo effect in a metallic base where the condensed chargons induce the local CB condensation at the impurity~\cite{Florens3}, the magnetic impurity on $U\left(1\right)$ QSL with SFS has the impurity charged bosonic state mixed with a fraction of chargon states in the QSL.

With both the channels of the SAF spinon coupling and the CB chargon coupling analyzed, the electronic Matsubara Green's function for the magnetic impurity based on the QSL is ready to be obtained from the convolution in Eq. \ref{Convolution}, which yields the form
\begin{align}\nonumber
G^0_{d, \sigma}\left(i\omega_n\right)\approx&\frac{1}{2}\left[\frac{1}{i\omega_n-\epsilon_0+\frac{U}{2}}+\frac{1}{i\omega_n-\epsilon_0-\frac{U}{2}}\right.\\
&\left.+Z\sum_{\bm{k}}\left(\frac{1}{i\omega_n+\epsilon_{\bm{k}}}+\frac{1}{i\omega_n-\epsilon_{\bm{k}}}\right)\right].
\end{align}
The impurity bare electronic spectral function is thus derived to be $\rho^0_{d, \sigma}\left(\omega\right)\approx\frac{1}{2}\left[\delta\left(\omega-\epsilon_0+\frac{U}{2}\right)+\delta\left(\omega-\epsilon_0-\frac{U}{2}\right)+Z|\rho_X\left(\omega\right)|\right]$ with $\rho_{X}\left(\omega\right)=-\frac{1}{\pi}\textrm{Im}G_X\left(i\nu_n\rightarrow\omega+i0^+, \bm{R}, \bm{R}\right)$. The calculations of $G^0_{d, \sigma}\left(i\omega_n\right)$ and $\rho^0_{d, \sigma}\left(\omega\right)$ through convolution are carried out in the spectral representation, which can be found in Eq. \ref{Covolut_G_d} and \ref{Convolut_rho_d} in Appendix \ref{App-D}. In addition to the atomic energy levels induced peaks at $\omega=\epsilon_0\pm\frac{U}{2}$, the impurity bare electronic spectral function $\rho^0_{d, \sigma}\left(\omega\right)$ now involves the chargon DOS $|\rho_X\left(\omega\right)|$ with a spectral weight $Z$, as the dashed spectrum shows in Fig. \ref{figure2} (c). The peak in the spectrum comes from van Hove singularity in the tight binding chargon band that we assumed. These impurity states at the energy of chargon bands take a fraction of QSL electronic states transferred through the spinon Kondo resonance in the SAF spinon channel and the hybridization in the CB chargon channel. The magnetic impurity on $U\left(1\right)$ QSL with SFS therefore acts as a local probe of the chargon DOS in the QSL.

\section{Full Mean Field Calculation for a Magnetic Impurity in a QSL}\label{IV}
To verify the above analysis of a magnetic impurity weakly coupled to a $U\left(1\right)$ QSL with SFS, the self-consistent equations Eq. \ref{SF_Eq1} - \ref{SF_Eq4} are numerically solved in the spectral representation (see Appendix \ref{App-D} for the detailed self-consistent integral equations in the spectral reprsentation). Concerning about the recent two dimensional $U\left(1\right)$ QSL candidates monolayer TaSe$_2$~\cite{Crommie1, Crommie2, Crommie4}, which is in triangular lattice, we take the spinon band dispersion $\xi_{\bm{k}}$ and the chargon band dispersion $\epsilon_{\bm{k}}$ to be $\xi_{\bm{k}}=2t_f\left(2\cos\frac{1}{2}k_xa\cos\frac{\sqrt{3}}{2}k_ya+\cos k_xa\right)-\mu_f$ and $\epsilon_{\bm{k}}=2t_X\left(2\cos\frac{1}{2}k_xa\cos\frac{\sqrt{3}}{2}k_ya+\cos k_xa-3\right)+\Delta$ respectvely. Here $a$ is the lattice constant, $t_f=0.05$eV is the spinon hopping, $t_X=-0.03$eV is the chargon hopping, $\mu_f=-0.04$eV is the spinon chemical potential that makes  the spinon band half-filled, and $\Delta=0.25$eV is the Mott gap size to match the recent STS measurement on TaSe$_2$~\cite{Crommie4}. The spinon spectral function $\rho_{f, \sigma}\left(\omega\right)=-\frac{1}{\pi}\textrm{Im}\sum_{\bm{k}}\frac{1}{\omega+i0^+-\xi_{\bm{k}}}$, chargon spectral function $\rho_{X}\left(\omega\right)=-\frac{1}{\pi}\textrm{Im}\sum_{\bm{k}}\left(\frac{1}{\omega+i0^++\epsilon_{\bm{k}}}-\frac{1}{\omega+i0^+-\epsilon_{\bm{k}}}\right)$ and the convolution generated electronic spectral function $\rho_{c, \sigma}\left(\omega\right)$ for the $U\left(1\right)$ QSL in triangular lattice have been calculated and shown in Fig. \ref{figureS1}. At the magnetic impurity, we set the impurity onsite energy to be $\epsilon_0=0$ eV, which corresponds to the symmetric Anderson impurity model. The Coulomb repulsion at the impurity is considered to be $U=3$eV, which is much larger than the Mott gap $\Delta$, so the atomic energy levels do not have energy overlap with the Hubbard bands in QSL. The full numerical solutions to the self-consistent equations in Eq. \ref{SF_Eq1} - \ref{SF_Eq4} yield the SAF spinon coupling $w$ and CB chargon coupling $u$ as a function of the electronic coupling $V_{\bm{R}}$, as is shown in Fig. \ref{figure1} (b). The $w$ and $u$ are extremely small at weak coupling $V_{\bm{R}}$, and then become significantly nonzero as $V_{\bm{R}}$ increases. With a given electronic coupling $V_{\bm{R}}$, the nontrivial solutions of $w$ and $u$ from Eq. \ref{SF_Eq1} - \ref{SF_Eq4} are always guaranteed as long as the onsite interaction $U$ is sufficiently large to induce a local magnetic moment formed at the impurity.

With the self-consistent equations generated mean field parameters $w$, $u$, $h$ and $\lambda$, the SAF spectral function $\rho_{a, \sigma}\left(\omega\right)=-\frac{1}{\pi}\textrm{Im}G_{a, \sigma}\left(i\omega_n\rightarrow\omega+i0^+\right)$ and the CB spectral function $\rho_{X_d}\left(\omega\right)=-\frac{1}{\pi}\textrm{Im}G_{X_d}\left(i\nu_n\rightarrow\omega+i0^+\right)$ are numerically calculated and shown in Fig. \ref{figure2} (d) and (e). Consistent with our analysis for a magnetic impurity weakly coupled to a $U\left(1\right)$ QSL with SFS, the SAF spectral function $\rho_{a, \sigma}\left(\omega\right)$ has the spinon Kondo resonance at the Fermi energy, and the CB spectral function $\rho_{X_d}\left(\omega\right)$  in the QSL chargon band energy region $-\Lambda_X<\omega<\Lambda_X$ replicates the QSL chargon spectral function $\rho_X\left(\omega\right)$ with a spectral weight $Z$. The original spectral feature of the CB at $\omega=\pm\frac{1}{2}U$ appears as well in the presence of coupling $u$. Convolving the SAF Green's function $G_{a, \sigma}\left(i\omega_n\rightarrow\omega+i0^+\right)$ with $G_{X_d}\left(i\nu_n\rightarrow\omega+i0^+\right)$ then yields the bare electronic spectral function $\rho^0_{d, \sigma}\left(\omega\right)$ for the magnetic impurity based on the QSL of triangular lattice, as is shown in Fig. \ref{figure2} (f). In addition to the peaks from the two atomic energy levels at $\omega=\pm\frac{1}{2}U$, $\rho^0_{d, \sigma}\left(\omega\right)$ at $-\Lambda_X<\omega<\Lambda_X$ becomes nonzero and resembles the QSL chargon DOS $|\rho_X\left(\omega\right)|$ with a spectral weight. The two peaks shown in $\rho^0_{d, \sigma}\left(\omega\right)$ at $-\Lambda_X<\omega<\Lambda_X$ are from the Van Hove singularities of the QSL chargon bands. At the Hubbard band edge energies $\omega=\pm\Delta$, the impurity bare electronic spectral function $\rho^0_{d, \sigma}\left(\omega\right)$ has a step that is inherited from the QSL chargon DOS $|\rho_X\left(\omega\right)|$. The bare impurity spectral function $\rho_{d, \sigma}^0\left(\omega\right)$ from the full numerical self-consistent calculations again confirm our analysis that the magnetic impurity on $U\left(1\right)$ QSL with SFS locally measures the chargon DOS in the QSL.

\begin{figure*}
\centering
\includegraphics[width=6.5in]{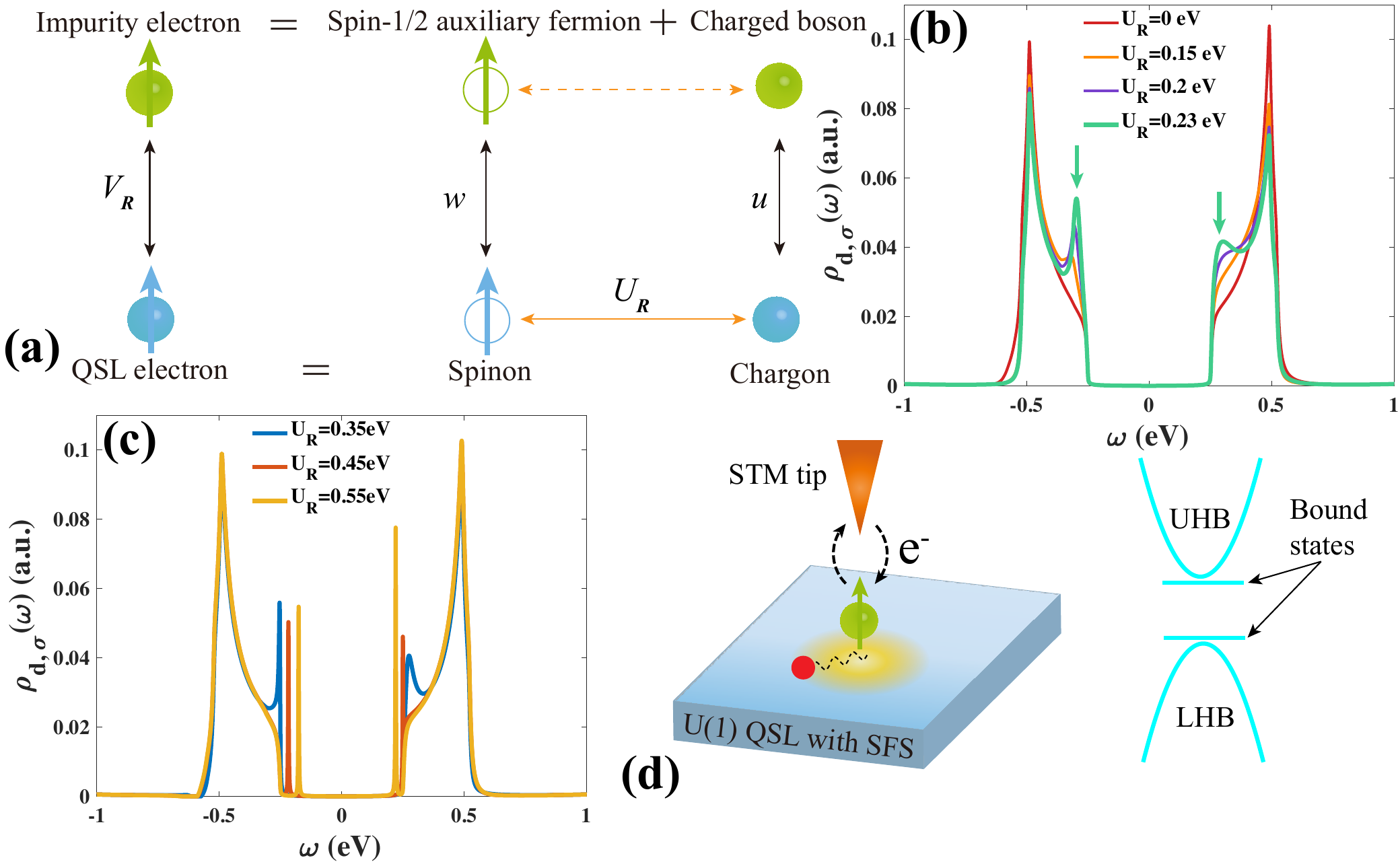}
\caption{(a) The schematic diagram to show the spinon chargon binding interaction incorporated in the slave rotor mean field description. The impurity electronic state is affected by the spinon chargon binding interaction in the QSL. (b) The impurity electronic spectral function $\rho_{d, \sigma}\left(\omega\right)$. As the spinon chargon binding interaction in the QSL increases from 0, a pair of resonance peaks emerge at the Hubbard band edge energies as indicated by the arrows. The original atomic energy levels at $\omega=\pm\frac{U}{2}$ lie outside the range of energies we plot. (c) The in-gap bound state peaks in $\rho_{d, \sigma}\left(\omega\right)$. As the binding interaction further increases, the band edge resonance peaks move inside the Mott gap and evolve to be in-gap bound state peaks. (d) A schematic plot of an STM setup to measure the density of states spectroscopy of the magnetic impurity. The filled red circle with a wavy line connected to the magnetic impurity denotes the hole or electronic state bound to the magnetic impurity. In the negative bias of $\omega\approx-\Delta$, the hole state bound to the magnetic impurity is excited so that an in-gap bound state peak emerges above the top of the LHB. Similarly, a positive bias of $\omega\approx\Delta$ injects an electron into the magnetic impurity and excites the electronic state bound to the magnetic impurity, generating an electronic bound state peak below the UHB bottom.}\label{figure3}
\end{figure*}

\section{Band Edge Resonance Peaks and In-gap Bound States from Spinon Chargon Binding Interaction}\label{V}
The step edges of $\rho^0_{d, \sigma}\left(\omega\right)$ at the Hubbard band edge energies $\omega=\pm\Delta$ make the impurity bare electronic states susceptible to the residue spinon chargon interaction in the QSL host. In the $U\left(1\right)$ QSL with SFS, the spinons and chargons both couple to the emergent $U\left(1\right)$ gauge field, so the gauge field fluctuations in turn modify the QSL electronic states and further affect the impurity electronic state that couples to electrons in the QSL host. At the Hubbard band edge energies $\omega=\pm\Delta$, the transverse components of the gauge field fluctuations are neglected here due to the small current current correlation at band edges, so the longitudinal component of the gauge field fluctuations plays the dominant role~\cite{Tang}. Since the spinons and chargons carry the opposite gauge charge, the longitudinal component of gauge field fluctuations yields a binding between a spinon and a chargon, which is the spinon chargon binding interaction~\cite{Lee4, Lee5}. In the presence of spinon chargon binding interaction at the QSL, a spinon tends to get bound with a chargon to form a physical electronic state. At the magnetic impurity, the SAF has spin exchange with the spinon Kondo cloud and the CB gets hybridized with the nearby chargons, so the spinon chargon binding interaction in the QSL host affects the impurity electronic state, as is schematically shown in Fig. \ref{figure3} (a).

To take into account the spinon chargon binding interaction in the QSL, we notice that the chargons in the QSL are relativistic~\cite{Lee3} and contain both the holons and doublons: $X_{\bm{r}}=a_{\bm{r}}+b^\dagger_{\bm{r}}$, $X^\dagger_{\bm{r}}=a^\dagger_{\bm{r}}+b_{\bm{r}}$, with the holon annihilation (creation) operator $a^{\left(\dagger\right)}_{\bm{r}}$ and the doublon annihilation (creation) operator $b^{\left(\dagger\right)}_{\bm{r}}$. Assuming a noncompact U(1) gauge field, the fluctuations of the longitudinal component give rise to the spinon chargon binding interaction term:
\begin{align}\label{Act_prime}
S'=&\int_0^\beta\int\sum_\sigma U_{\bm{r}}f^\dagger_{\sigma, \bm{r}}f_{\sigma, \bm{r}}\left(a^\dagger_{\bm{r}}a_{\bm{r}}-b^\dagger_{\bm{r}}b_{\bm{r}}\right)d\bm{r}d\tau,
\end{align}
where $U_{\bm{r}}>0$ is the spinon chargon binding interaction strength. The detailed derivation of Eq. \ref{Act_prime} can be found in Appendix \ref{App-E}. Since the longitudinal interaction is screened by the spinon Fermi sea, we approximate the interaction as short range and on site only. In Appendix \ref{App-E} the binding interaction strength has been estimated to be around half of the spinon band width: $U_{\bm{r}}\approx\Lambda_f$. The interaction between a spinon and a doublon in $S'$ is attractive so a spinon tends to get bound with a doublon to form an electronic bound state. On the other hand, the repulsion between a spinon and a holon in $S'$ manifests the attraction between a spinon hole and a holon, so a spinon hole tends to bind a holon to form a hole bound state. 
Around a magnetic impurity, the spinon Kondo cloud has both spinon excitations and spinon hole excitations, so in principle both an electronic state and a hole state bound to the magnetic impurity can be induced by a sufficiently large binding interaction.

In the presence of the spinon chargon binding interaction, the action for a magnetic impurity embedded in a $U\left(1\right)$ QSL with SFS becomes $S=S_0+S'$. At the impurity, the charged bosonic operator can also be written in terms of the holon and doublon operators as $X_d=a_d+b_d^\dagger$, $X_d^\dagger=a_d^\dagger+b_d$, so the impurity electronic Matsubara Green's function $G_{d, \sigma}\left(i\omega_n\right)$ is to calculate the thermal average from $S$:
\begin{align}\nonumber
G_{d, \sigma}\left(i\omega_n\right)=&-\int_0^\beta\left\langle a_\sigma\left(\tau\right)a^\dagger_{\sigma}\left(0\right)\right\rangle\left[\left\langle a^\dagger_d\left(\tau\right)a_d\left(0\right) \right\rangle\right.\\
&\left.+\left\langle b_d\left(\tau\right)b_d^\dagger\left(0\right) \right\rangle\right]d\tau.
\end{align}
Due to the spinon chargon binding interaction in the QSL, a doublon tends to get bound with a spinon to form an electronic state and a holon tends to get bound with a spinon hole to form a hole state. It is known that the CB $X_d=a_d+b^\dagger_d$ at the impurity couples to the chargons in the QSL through the mean field parameter $u$, and the SAF $a_\sigma$ at the impurity couples to the spinons in the QSL through the mean field parameter $w$, so the impurity electronic Matsubara Green's function can be calculated through the random phase approximation. The resulting impurity electronic Matsubara Green's function $G_{d, \sigma}\left(i\omega_n\right)$ takes the form
\begin{widetext}
\begin{align}\label{G_d_sigma}
G_{d, \sigma}\left(i\omega_n\right)=G^0_{d, \sigma}\left(i\omega_n\right)+\sum_{i, j=1, 2}\left\{\hat{T}_\sigma\left(i\omega_n\right)U_{\bm{R}}\sigma_z\left[1-\hat{G}_{c, \sigma}\left(i\omega_n\right)U_{\bm{R}}\sigma_z\right]^{-1}\hat{T}_\sigma\left(i\omega_n\right)\right\}_{ij}.
\end{align}
The detail derivation of $G_{d, \sigma}\left(i\omega_n\right)$ can be found in Appendix \ref{App-F}. Here $\sigma_z$ is the $z$ component of Pauli matrix acting on the holon and doublon space, and $\hat{T}_\sigma\left(i\omega_n\right)$, $\hat{G}_{c, \sigma}\left(i\omega_n\right)$ are both the $2\times2$ correlators
\begin{align}
\hat{T}_\sigma\left(i\omega_n\right)=&-\int_0^\beta\left\langle a_\sigma\left(\tau\right)f^\dagger_{\sigma, \bm{R}}\left(0\right) \right\rangle_0\left\langle \begin{pmatrix}
a^\dagger_d\left(\tau\right)a_{\bm{R}}\left(0\right) & a^\dagger_d\left(\tau\right)b^\dagger_{\bm{R}}\left(0\right) \\
b_d\left(\tau\right)a_{\bm{R}}\left(0\right) & b_d\left(\tau\right)b^\dagger_{\bm{R}}\left(0\right)
\end{pmatrix}\right\rangle_0d\tau,\\
\hat{G}_{c, \sigma}\left(i\omega_n\right)=&-\int_0^\beta\left\langle f_{\sigma, \bm{R}}\left(\tau\right)f^\dagger_{\sigma, \bm{R}}\left(0\right)\right\rangle_0\left\langle \begin{pmatrix}
a^\dagger_{\bm{R}}\left(\tau\right)a_{\bm{R}}\left(0\right) & a^\dagger_{\bm{R}}\left(\tau\right)b^\dagger_{\bm{R}}\left(0\right) \\
b_{\bm{R}}\left(\tau\right)a_{\bm{R}}\left(0\right) & b_{\bm{R}}\left(\tau\right)b^\dagger_{\bm{R}}\left(0\right)
\end{pmatrix} \right\rangle_0d\tau,
\end{align}
\end{widetext}
with $\left\langle\dots\right\rangle_0$ being the thermal average calculated from $S_0$ in Eq. \ref{Act_tot}. The more detail form of $\hat{T}_\sigma\left(i\omega_n\right)$ and $\hat{G}_{c, \sigma}\left(i\omega_n\right)$ have been derived in Appendix \ref{App-F}. The spinon chargon binding interaction $U_{\bm{R}}$ in the QSL modifies the impurity electronic Green's function through the second term in Eq. \ref{G_d_sigma}.

As the spinon chargon binding interaction increases from 0, it induces the spectral weight to transfer from the bulk Hubbard bands to the band edges. At $U_{\bm{R}}=0.23$ eV, which is around half of the spinon band width we considered (see Fig. \ref{figureS1} in Appendix \ref{App-E}), a pair of band edge resonance peaks arise as seen in Fig. \ref{figure3} (b). The band edge resonance peaks indicated by the arrows in Fig. \ref{figure3} (b) are still located inside the Hubbard bands, so the band edge resonance peaks have the intrinsic width that is determined by the spinons, chargons, and their coupling to the impurity electronic state. Importantly, the band edge resonance peaks appear only at the magnetic impurity but do not show up in the pristine QSL given the same binding interaction (see Fig, \ref{figureS2} in Appendix \ref{App-F} for comparison). Such band edge resonance peaks at the Hubbard band edges can be regarded as the precursors of in-gap bound states.

As the spinon chargon binding interaction further increases, the band edge resonance peaks gradually move inside the Mott gap and develop to be in-gap peaks of bound states as is shown in Fig. \ref{figure3} (c). The in-gap bound states arise from the new poles of $G_{d, \sigma}\left(i\omega_n\rightarrow\omega+i0^+\right)$ that come from $U_{\bm{R}}\sigma_z\left[1-\hat{G}_{c, \sigma}\left(i\omega_n\rightarrow\omega+i0^+\right)U_{\bm{R}}\sigma_z\right]^{-1}$ at a sufficiently large binding $U_{\bm{R}}$, so an electronic bound state and a hole bound state appear below the UHB bottom and above the LHB top respectively as shown in Fig. \ref{figure3} (c) and (d). The in-gap peaks of the bound states have no intrinsic width so we introduce an extrinsic broadening of 5 meV to make them better visualized in Fig. \ref{figure3} (c).

Physically, the band edge resonance peaks in the magnetic impurity spectral function $\rho_{d, \sigma}\left(\omega\right)$ near $\omega\approx\pm\Delta$ and the in-gap peaks of the bound states can be interpreted as the joint effect of the spinon Kondo screening cloud and the spinon chargon binding interaction around the magnetic impurity. At the top of the LHB, a spinon hole excitation in the spinon Kondo cloud attracts a holon to form a hole excitation. The hole excitation is manifested as a band edge resonance peak at a moderate binding interaction and it evolves into an in-gap hole bound state as the binding interaction further increases. In analogy to an acceptor impurity embedded in a semiconductor, the hole excitation and its descendent hole bound state around the magnetic impurity appear at the top of the LHB as seen in Fig. \ref{figure3} (b)-(d). At the bottom of the UHB, the attraction between a spinon in the spinon Kondo cloud and a doublon gives rise to an electronic excitation, which also shows up as a band edge resonance peak in $\rho_{d, \sigma}\left(\omega\right)$. As the binding interaction increases, the band edge resonance peak moves down into the Mott gap and develops to be an in-gap electronic bound state peak. Similar to a donor impurity in a semiconductor, the electronic excitation and the descendent in-gap electronic bound state around the magnetic impurity form at the bottom of the UHB as shown in Fig. \ref{figure3} (b)-(d). When a scanning tunneling microscope (STM) tip is put on top of the magnetic impurity as seen in Fig. \ref{figure3} (d), applying a negative bias with $\omega\approx-\Delta$ extracts an electron out of the QSL and creates a hole near the magnetic impurity. Depending on the binding interaction strength, either a band edge resonance peak at the LHB top or a hole bound state peak above the LHB top appears. On the other hand, applying a positive bias with $\omega\approx\Delta$ injects an electron into the QSL through the magnetic impurity, which induces either a band edge resonance peak at the UHB bottom or an electronic bound state peak below the UHB bottom, depending on the binding interaction strength. Since the spinon Kondo screening cloud has both the spinon hole excitations attractive to the holons and the spinon excitations attractive to the doublons, the magnetic impurity screened by the spinon cloud acts as an impurity of acceptor and donor type simultaneously, generating a peak located at the top of the UHB and one at the bottom of the LHB. Such spectral feature induced by a magnetic impurity embedded in a Mott insulator therefore indicates that the host Mott insulator is a $U\left(1\right)$ QSL with SFS. Since the effective attraction comes from the longitudinal component of gauge field fluctuations, the peaks near $\omega\approx\pm\Delta$ in the spectroscopy of magnetic impurity manifest the gauge field fluctuations in the $U\left(1\right)$ QSL with SFS. 

\section{Conclusion and discussions}\label{VI}
In the above sections, applying the slave rotor mean field theory and incorporating the longitudinal component of gauge field fluctuations, the electronic spectral function of a magnetic impurity embedded in a $U\left(1\right)$ QSL with SFS was calculated. At the mean field level where the gauge field fluctuations are neglected in the QSL, the magnetic impurity has the spinon Kondo screening cloud around and locally measures the chargon DOS in the QSL. When the gauge field fluctuations are included in the QSL, the resulting spinon chargon binding interaction in the spinon Kondo cloud induces a pile-up of spectral weight at the Hubbard band edges, which shows up as band edge resonance peaks in the spectroscopy of the magnetic impurity. The band edge resonance peaks appear due to the joint effect of the spinon Kondo cloud and the gauge field fluctuations at the QSL sites around the magnetic impurity. As a result, in the spectroscopy of a magnetic impurity, a pair of resonance peaks emerging at $\omega\approx\pm\Delta$ of the Hubbard band edges provide the strong evidence that the host Mott insulator is a $U\left(1\right)$ QSL with SFS.

Recently, scanning tunneling spectroscopy measurements have been carried out on a Co adatom embedded in a monolayer TaSe$_2$~\cite{Crommie4}, which is a QSL candidate. The spectroscopy measurements on top of the Co adatom exhibit a resonance peak at both the LHB top and UHB bottom, while a contrast measurement on an Au impurity shows no resonance peaks. As the Co atom carries a local magnetic moment, it matches our theory of a magnetic impurity embedded on a $U\left(1\right)$ QSL with SFS. Based on our theory, the observation of the resonance peaks in the spectroscopy of a Co adatom strongly suggests that the monolayer Mott insulator TaSe$_2$ has the spinon Fermi surface and emergent $U\left(1\right)$ gauge field.

Our establishment of the spectral function for magnetic impurity on $U\left(1\right)$ QSL with SFS suggests that depositing magnetic impurity on a Mott insulator and carrying out spectroscopy measurements on the impurity can serve as a diagnosis for the QSL state. As recently new two dimensional electronic correlation induced insulators emerge with Moir\'e flat bands~\cite{Pablo1, Pablo2, Balents2}, the magnetic impurity deposit followed by spectroscopy measurements is thus expected to identify more candidates for $U\left(1\right)$ QSL with SFS.

\section*{ACKNOWLEDGMENTS}
The authors acknowledge helpful discussions with Yi Chen, Michael F. Crommie and Serge Florens. W.-Y. He acknowledges the start-up grant of ShanghaiTech University. P. A. L. acknowledges support by DOE office of Basic Sciences grant number DE-FG02-03-ER46076.

\appendix
\setcounter{figure}{0}
\renewcommand{\thefigure}{A\arabic{figure}}

\section{SPECTRAL REPRESENTATION OF GREEN'S FUNCTION} \label{App-A}
We first review about the spectral representation of a Green's function. For a Green's function $\mathcal{G}\left(z\right)$, its spectral representation reads
\begin{align}
\mathcal{G}\left(z\right)=&-\int_{-\infty}^\infty\frac{\textrm{Im}\mathcal{G}\left(\omega'+i0^+\right)}{z-\omega}\frac{d\omega'}{\pi}.
\end{align}
Here $z$ is complex. When $z$ takes $z=i\omega_n$ or $z=i\nu_n$, the Green's function $\mathcal{G}\left(z\right)$ is a Matsubara Green's function. When $z$ takes $z=\omega+i0^+$, the Green's function $\mathcal{G}\left(z\right)$ is a retarded Green's function. The spectral representation of Green's function is widely used in our calculations.

For the $U\left(1\right)$ QSL with SFS, its onsite spinon Matsubara Green's function takes the form
\begin{align}\nonumber
G_{f, \sigma}\left(i\omega_n, \bm{R}, \bm{R}\right)=&-\left\langle f_{\sigma, \bm{R}}\left(i\omega_n\right)f^\dagger_{\sigma, \bm{R}}\left(i\omega_n\right)\right\rangle_0e^{i\left(\bm{k}-\bm{k}'\right)\cdot\bm{R}}\\
=&\frac{1}{\Omega}\sum_{\bm{k}}\frac{1}{i\omega_n-\xi_{k}},
\end{align}
with $\Omega$ the volume of the system. The spinon Matsubara Green's function can also be expressed in the spectral representation as
\begin{align}
G_{f, \sigma}\left(i\omega_n, \bm{R}, \bm{R}\right)=&\int_{-\infty}^\infty\frac{\rho_{f, \sigma}\left(\omega'\right)}{i\omega_n-\omega'}d\omega',
\end{align}
with the spinon spectral function $\rho_{f, \sigma}\left(\omega'\right)=-\frac{1}{\pi}\textrm{Im}G^{\textrm{R}}_{f, \sigma}\left(\omega'\right)$ and $G^{\textrm{R}}_{f, \sigma}\left(\omega'\right)=G_{f, \sigma}\left(i\omega_n\rightarrow\omega'+i0^+\right)$. Similarly, the onsite chargon Green's function can be written as
\begin{align}\nonumber
G_X\left(i\nu_n, \bm{R}, \bm{R}\right)=&\left\langle X_{\bm{R}}\left(i\nu_n\right)X^\dagger_{\bm{R}}\left(i\nu_n\right)\right\rangle_0\\
=&\frac{1}{\Omega}\sum_{\bm{k}}\left(\frac{1}{i\nu_n+\epsilon_{\bm{k}}}-\frac{1}{i\nu_n-\epsilon_{\bm{k}}}\right),
\end{align}
and its spectral representation reads
\begin{align}
G_{X}\left(i\nu_n, \bm{R}, \bm{R}\right)=&\int_{-\infty}^\infty\frac{\rho_X\left(\omega'\right)}{i\nu_n-\omega'}d\omega',
\end{align}
with the chargon spectral function $\rho_X\left(\omega'\right)=-\frac{1}{\pi}\textrm{Im}G^{\textrm{R}}_{X}\left(\omega'\right)$ and $G^{\textrm{R}}_X\left(\omega'\right)=G_X\left(i\nu_n\rightarrow\omega'+i0^+\right)$.

\section{DERIVATION OF THE TWO APPROXIMATED ALGEBRA EQUATIONS}\label{App-B}
In the isolated magnetic impurity with zero coupling, the SAF has the bare Matsubara Green's function $G^0_{a, \sigma}\left(i\omega_n\right)=\frac{1}{i\omega_n}$, and the CB has the bare Matsubara Green's function $G^0_{X_d}\left(i\nu_n\right)=\frac{1}{i\nu_n+\epsilon_0+\frac{U}{2}}-\frac{1}{i\nu_n+\epsilon_0-\frac{U}{2}}$~\cite{Florens2}. 

In the magnetic impurity weakly coupled to the $U\left(1\right)$ QSL with SFS, by taking the approximated Lagrangian multipliers to be $h=\epsilon_0$, $\lambda=\frac{U}{4}-\frac{h^2}{U}$, the Matsubara frequency summation in Eq. \ref{SF_Eq1} can be calculated as
\begin{align}\nonumber
&\frac{1}{\beta}\sum_{\omega_n}G_{a, \sigma}\left(i\omega_n\right)G_{f, \sigma}\left(i\omega_n, \bm{R}, \bm{R}\right)\\\nonumber
=&-\frac{1}{\pi}\int_{-\infty}^\infty\textrm{Im}\frac{G_1^{\textrm{R}}\left(\omega\right)}{1-w^2G_1^{\textrm{R}}\left(\omega\right)}n_{\textrm{F}}\left(\omega\right)d\omega\\
\approx&-\frac{1}{\pi}\int_{-\infty}^0\frac{\textrm{Im}G_1^{\textrm{R}}\left(\omega\right)}{1+\left[w^2\textrm{Im}G_1^{\textrm{R}}\left(\omega\right)\right]^2}d\omega,
\end{align}
with 
\begin{align}\nonumber
G^{\textrm{R}}_1\left(\omega\right)=&G^0_{a, \sigma}\left(i\omega_n\rightarrow\omega+i0^+\right)G_{f, \sigma}\left(i\omega_n\rightarrow\omega+i0^+, \bm{R}, \bm{R}\right)\\
=&\frac{1}{\omega+i0^+}\left[\int_{-\infty}^\infty\frac{\rho_{f, \sigma}\left(\omega'\right)}{\omega-\omega'}d\omega'-i\pi\rho_{f, \sigma}\left(\omega\right)\right],
\end{align}
and the Fermi distribution function $n_{\textrm{F}}\left(\omega\right)=\frac{1}{e^{\beta\omega}+1}$. For simplicity, the spinon DOS is assumed to be constant $\rho_{f, \sigma}\left(\omega\right)=\frac{1}{2\Lambda_f}\Theta\left(\Lambda^2_f-\omega^2\right)$ and then self-consistent equation in Eq. \ref{SF_Eq1} is simplified to be
\begin{align}\label{Algebra1}
u\approx&-2wV_{\bm{R}}\int_{-\Lambda_f}^0\frac{\omega}{\omega^2+\Gamma_{\textrm{eff}}^2}\frac{1}{2\Lambda_f}d\omega=-\frac{wV_{\bm{R}}}{2\Lambda_f}\ln\frac{\Gamma^2_{\textrm{eff}}}{\Lambda_f^2+\Gamma^2_{\textrm{eff}}},
\end{align}
with $\Gamma_{\textrm{eff}}=\frac{\pi w^2}{2\Lambda_f}$. In Eq. \ref{SF_Eq2}, the Matsubara frequency summation can be calculated as
\begin{align}\nonumber
&\frac{1}{\beta}\sum_{\nu_n}G_{X_d}\left(i\nu_n\right)G_X\left(i\nu_n, \bm{R}, \bm{R}\right)\\\nonumber
=&\frac{1}{\pi}\int_{-\infty}^\infty\textrm{Im}\frac{G_2^{\textrm{R}}\left(\omega\right)}{1-u^2G_2^{\textrm{R}}\left(\omega\right)}n_{\textrm{B}}\left(\omega\right)d\omega\\
\approx&-\frac{1}{\pi}\int_{-\infty}^\infty\frac{\textrm{Im}G_2^{\textrm{R}}\left(\omega\right)}{1+\left[u^2\textrm{Im}G_2^{\textrm{R}}\left(\omega\right)\right]^2}d\omega,
\end{align}
with
\begin{align}\nonumber
G_2^{\textrm{R}}\left(\omega\right)=&G^0_{X_d}\left(i\nu_n\rightarrow\omega+i0^+\right)G_X\left(i\nu_n\rightarrow\omega+i0^+, \bm{R}, \bm{R}\right)\\
\approx&\frac{4U}{U^2-4\epsilon_0^2}\left[\int_{-\infty}^\infty\frac{\rho_X\left(\omega'\right)}{\omega-\omega'}d\omega'-i\pi\rho_X\left(\omega\right)\right],
\end{align}
and the Bose distribution function $n_{\textrm{B}}\left(\omega\right)=\frac{1}{e^{\beta\omega}-1}$. For simplicity, the chargon DOS is assumed to be constant: $\rho_X\left(\omega\right)=\frac{1}{\Lambda_X-\Delta}$ for $-\Lambda_X<\omega<-\Delta$ and $\rho_X\left(\omega\right)=-\frac{1}{\Lambda_X-\Delta}$ for $\Delta<\omega<\Lambda_X$. Then the self-consistent equation in Eq. \ref{SF_Eq2} is simplified to be
\begin{align}\label{Algebra2}
w=&uV_{\bm{R}}\int_{-\Lambda_X}^{-\Delta}\frac{\frac{4U}{U^2-4\epsilon_0^2}\frac{1}{\Lambda_X-\Delta}}{1+\left[\frac{4\pi u^2U}{\left(U^2-4\epsilon_0^2\right)\left(\Lambda_X-\Delta\right)}\right]^2}d\omega\approx\frac{4uUV_{\bm{R}}}{U^2-4\epsilon_0^2}.
\end{align}
In a magnetic impurity weakly coupled to a $U\left(1\right)$ QSL with SFS, the mean field self-consistent equations in Eq. \ref{SF_Eq1} and Eq. \ref{SF_Eq2} are then approximated by the two algebra equations in Eq. \ref{Algebra1} and Eq. \ref{Algebra2} respectively.

\section{DERIVATION OF THE SPECTRAL WEIGHT}\label{App-C}
In the presence of mean field coupling $u$, the CB Matsubara Green's function in Eq. \ref{Green_boson} can be approximated by the Dyson equation as
\begin{align}\nonumber
G_{X_d}\left(i\nu_n\right)=&\frac{G^0_{X_d}\left(i\nu_n\right)}{1-u^2G^0_{X_d}\left(i\nu_n\right)G_X\left(i\nu_n, \bm{R}, \bm{R}\right)}\\
\approx&G_{X_d}^0\left(i\nu_n\right)\left[1+u^2G_X\left(i\nu_n, \bm{R}, \bm{R}\right)G_{X_d}^0\left(i\nu_n\right)\right].
\end{align}
As the Coulomb repulsion $U$ is much larger than the low energy region we concern, the $G^0_{X_d}\left(i\nu_n\right)$ in the second term is approximated to be $G^0_{X_d}\left(i\nu_n\right)\approx \frac{4U}{U^2-4\epsilon_0^2}$ and it yields
\begin{align}
G_{X_d}\left(i\nu_n\right)\approx G^0_{X_d}\left(i\nu_n\right)+Z\sum_{\bm{k}}G_X\left(i\nu_n, \bm{R}, \bm{R}\right).
\end{align}
Here the $Z$ is the spectral weight that takes the value $Z=\frac{16u^2U^2}{\left(U^2-4\epsilon_0^2\right)^2}$. 

\begin{figure*}
\centering
\includegraphics[width=6.8in]{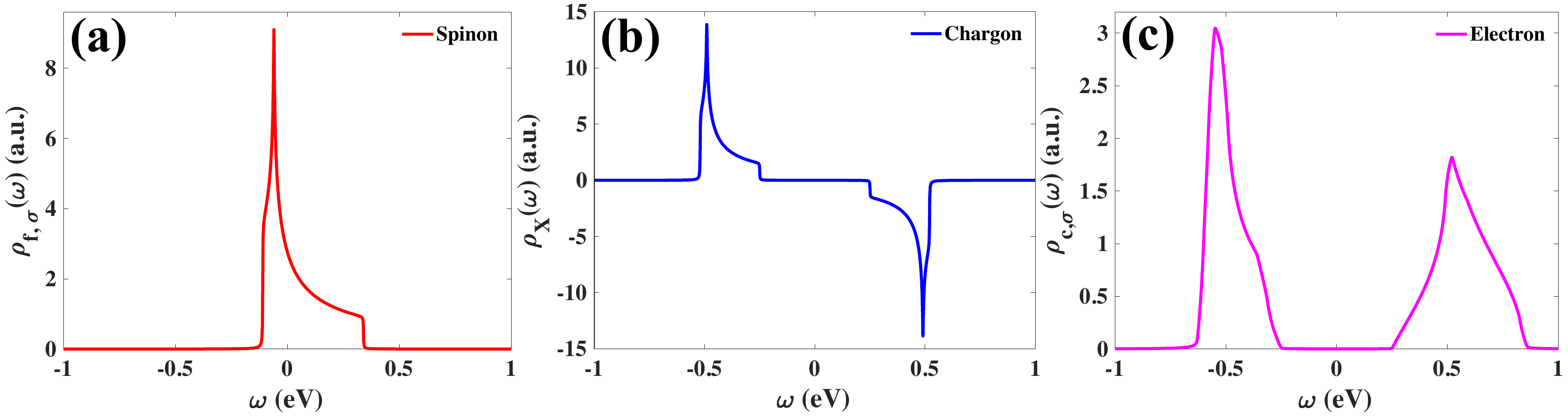}
\caption{(a) The spectral function $\rho_{f, \sigma}\left(\omega\right)$ for the itinerant spinons in the QSL. (b) The spectral function $\rho_X\left(\omega\right)$ for the QSL chargons. (c) The electronic spectral function $\rho_{c, \sigma}\left(\omega\right)$ for the $U\left(1\right)$ QSL with SFS. The QSL considered here is of triangular lattice.}\label{figureS1}
\end{figure*}

\section{SELF-CONSISTENT EQUATIONS IN THE SPECTRAL REPRESENTATION}\label{App-D}
By taking the spectral representation of Matsubara Green's functions, the self-consistent equations in Eq. \ref{SF_Eq1} - \ref{SF_Eq4} that involves the infinite summation series can all be written integral equations form as
\begin{widetext}
\begin{align}\label{Num_SF_Eq1}
u=&-2wV_{\bm{R}}\int_{-\infty}^\infty\left[\rho_{a, \sigma}\left(\omega\right)\textrm{Re}G^{\textrm{R}}_{f, \sigma}\left(\omega, \bm{R}, \bm{R}\right)+\textrm{Re}G_{a, \sigma}^{\textrm{R}}\left(\omega\right)\rho_{f, \sigma}\left(\omega\right)\right]n_{\textrm{F}}\left(\omega\right)d\omega,
\end{align}
\begin{align}\label{Num_SF_Eq2}
w=&-uV_{\bm{R}}\int_{-\infty}^\infty\left[\rho_{X_d}\left(\omega\right)\textrm{Re}G^{\textrm{R}}_X\left(\omega, \bm{R}, \bm{R}\right)+\rho_X\left(\omega\right)\textrm{Re}G^{\textrm{R}}_{X_d}\left(\omega\right)\right]n_{\textrm{B}}\left(\omega\right)d\omega,
\end{align}
\begin{align}\label{Num_SF_Eq3}
1=&-\int_{-\infty}^\infty\rho_{X_d}\left(\omega\right)n_{\textrm{B}}\left(\omega\right)d\omega,
\end{align}
\begin{align}\label{Num_SF_Eq4}
\int_{-\infty}^\infty\rho_{a, \sigma}\left(\omega\right)n_{\textrm{F}}\left(\omega\right)d\omega=&\frac{1}{2}-\frac{h}{U}+\frac{1}{2U}\int_{-\infty}^\infty\omega\rho_{X_d}\left(\omega\right)\coth\frac{1}{2}\beta\omega d\omega-\frac{1}{U\beta}\int_{-\infty}^\infty\rho_{X_d}\left(\omega\right)d\omega,
\end{align}
\end{widetext}
where the superscript $\textrm{R}$ means the retarded Green's function obtained by the analytic continuation $i\omega_n\rightarrow\omega+i0^+$, $i\nu_n\rightarrow\omega+i0^+$ in the corresponding Matsubara Green's function. 

In the numerical calculation, the spinon spectral function $\rho_{f, \sigma}\left(\omega\right)$ and the chargon spectral function $\rho_{X}\left(\omega\right)$ we considered is plotted in Fig. \ref{figureS1} (a) and (b) respectively. The resulting electronic spectral function $\rho_{c, \sigma}\left(\omega\right)$ can be obtained through the convolution in the spectral representation
\begin{align}\nonumber
\rho_{c, \sigma}\left(\omega\right)=&\int_{-\infty}^\infty\rho_{f, \sigma}\left(\omega+\omega'\right)\rho_X\left(\omega'\right)\left[n_{\textrm{F}}\left(-\omega-\omega'\right)\right.\\
&\left.+n_{\textrm{B}}\left(-\omega'\right)\right]d\omega'
\end{align}
and it is plotted in Fig. \ref{figureS1} (c).

With given spinon and chargon Green's function in the host system, the four integral equations can be numerically solved and yield the mean field parameters $w$, $u$, $\lambda$, $h$. In the spectral representation, the convolution of the impurity bare elecronic Matsubara Green's function is calculated as
\begin{align}\label{Covolut_G_d}
G_{d, \sigma}^0\left(i\omega_n\right)=&\frac{1}{\beta}\sum_{\nu_n}\sum_{-\infty}^{\infty}\frac{\rho_{a, \sigma}\left(\omega'\right)}{i\omega_n+i\nu_n-\omega'}d\omega'\int_{-\infty}^{\infty}\frac{\rho_{X_d}\left(\nu'\right)}{i\nu_n-\nu'}d\nu'.
\end{align}
By summing over the Matsubara frequency $\nu_n$, the retarded Green's function $G_{d, \sigma}^0\left(i\omega_n\rightarrow\omega+i0^+\right)$ can be obtained through the analytic continuation. The impurity bare electronic spectral function $\rho^0_{d, \sigma}\left(\omega\right)=-\frac{1}{\pi}\textrm{Im}G_{d, \sigma}^0\left(i\omega_n\rightarrow\omega+i0^+\right)$ then takes the form
\begin{align}\nonumber\label{Convolut_rho_d}
\rho^0_{d, \sigma}\left(\omega\right)=&\int_{-\infty}^\infty\rho_{a, \sigma}\left(\omega+\omega'\right)\rho_{X_d}\left(\omega'\right)\left[n_{\textrm{F}}\left(-\omega-\omega'\right)\right.\\
&\left.+n_{\textrm{B}}\left(-\omega'\right)\right]d\omega'.
\end{align}

\section{SPINON CHARGON BINDING INTERACTION FROM LONGITUDINAL GAUGE FIELD FLUCTUATIONS}\label{App-E}
The low energy effective action for the $U\left(1\right)$ QSL with SFS is~\cite{Senthil}
\begin{align}
S_{\textrm{SL}}=&S_f+S_\phi+S_{\textrm{M}},
\end{align}
with the spinon field action
\begin{align}\nonumber
S_f=&\int_0^\beta d\tau \int d\bm{r}\sum_\sigma\left[f^\dagger_{\sigma, \bm{r}}\left(\partial_\tau+iA_{0, \bm{r}}-\mu_f\right)f_{\sigma, \bm{r}}\right.\\
&\left.+\frac{\hbar^2}{2m_f}\left(\partial_{\bm{r}}+i\bm{A}_{\bm{r}}\right)f^\dagger_{\sigma, \bm{r}}\cdot\left(\partial_{\bm{r}}-i\bm{A}_{\bm{r}}\right)f_{\sigma, \bm{r}}\right],
\end{align}
the complex bosonic field action
\begin{align}\nonumber
S_\phi=&\int_0^\beta d\tau\int d\bm{r}\left[\left(\partial_\tau-iA_{0, \bm{r}}\right)\phi^\dagger_{\bm{r}}\left(\partial_\tau+iA_{0, \bm{r}}\right)\phi_{\bm{r}}\right.\\
&\left.+\hbar^2v^2_{\textrm{b}}\left(\partial_\tau+i\bm{A}_{\bm{r}}\right)\phi^\dagger_{\bm{r}}\left(\partial_\tau-i\bm{A}_{\bm{r}}\right)\phi_{\bm{r}}+\Delta^2\phi^\dagger_{\bm{r}}\phi_{\bm{r}}\right],
\end{align}
and the Maxwell term that controlls the gauge field fluctuations
\begin{align}
S_M=&\frac{1}{2g^2}\int_0^\beta d\tau\int d\bm{r}\left[\left(\nabla A_{0, \bm{r}}+\partial_\tau\bm{A}_{\bm{r}}\right)^2+\left(\nabla\times\bm{A}_{\bm{r}}\right)^2\right].
\end{align}
Here $m_f$ is the effective spinon mass, $v_{\textrm{b}}$ is the boson velocity, $\Delta$ is the chargon gap, $g$ is the bare gauge field coupling constant, and $A_{0, \bm{r}}$, $\bm{A}_{\bm{r}}$ are the temporary and spatial component of $U\left(1\right)$ gauge field respectively. The action $S_{\textrm{SL}}$ is invariant under the local $U\left(1\right)$ gauge transformation~\cite{Lee3}. At the mean field level where gauge fields are neglected, the spinon correlation function is $-\left\langle f_{\sigma, \bm{k}}\left(i\omega_n\right) f^\dagger_{\sigma, \bm{k}}\left(i\omega_n\right)\right\rangle_0=\frac{1}{i\omega_n-\xi_{\bm{k}}}$ with $\xi_{\bm{k}}=\frac{\hbar^2\bm{k}^2}{2m_f}-\mu_f$ the band dispersion in the continuum limit. The chargon operator becomes $X^{\left(\dagger\right)}_{\bm{k}}=\sqrt{2\epsilon_{\bm{k}}}\phi^{\left(\dagger\right)}_{\bm{k}}$ and the chargon correlation function reads $\left\langle X_{\bm{k}}\left(i\nu_n\right)X^\dagger_{\bm{k}}\left(i\nu_n\right)\right\rangle_0=\frac{1}{i\nu_n+\epsilon_{\bm{k}}}-\frac{1}{i\nu_n-\epsilon_{\bm{k}}}$ with $\epsilon_{\bm{k}}=\sqrt{\hbar^2v^2_{\textrm{b}}\bm{k}^2+\Delta^2}$ the chargon band dispersion in continuum limit. When the QSL is considered in a periodic lattice, tight binding band dispersions $\xi_{\bm{k}}$ and $\epsilon_{\bm{k}}$ can be constructed to replace those in the continuum limit.

In the Coulomb gauge $\nabla\cdot\bm{A}_{\bm{r}}=0$, as the group velocity at the Hubbard band edge energies is negligible, so the coupling of transverse gauge field to current is small and we drop the transverse gauge field fluctuations. The longitudinal component of gauge field $A_{0, \bm{r}}$ is kept in $S_f$, $S_\phi$ and $S_{\textrm{M}}$. The spinon field action $S_f$ describes the Schrodinger field, but the complex bosonic field $S_\phi$ describes the Klein Gordon field. The Klein Gordon field action $S_\phi$ can be rewritten in terms of the canonical momentum operators as
\begin{align}\nonumber\label{Klein_Gordon_Act}
S_\phi=&\int_0^\beta \int d\bm{r}\left[\Pi^\dagger_{\bm{r}}\partial_\tau\phi^\dagger_{\bm{r}}+\Pi_{\bm{r}}\partial_\tau\phi_{\bm{r}}-iA_{0, \bm{r}}\left(\Pi^\dagger_{\bm{r}}\phi^\dagger_{\bm{r}}-\Pi_{\bm{r}}\phi_{\bm{r}}\right)\right.\\
&\left.-\Pi^\dagger_{\bm{r}}\Pi_{\bm{r}}+\hbar^2v^2_{\textrm{b}}\partial_{\bm{r}}\phi^\dagger_{\bm{r}}\partial_{\bm{r}}\phi_{\bm{r}}+\Delta^2\phi^\dagger_{\bm{r}}\phi_{\bm{r}}\right],
\end{align}
with $\Pi^\dagger_{\bm{r}}=\left(\partial_\tau+iA_{0, \bm{r}}\right)\phi_{\bm{r}}$ and $\Pi_{\bm{r}}=\left(\partial_\tau-iA_{0, \bm{r}}\right)\phi^\dagger_{\bm{r}}$ the canonical momentum operators. We apply the Fourier transformation to Eq. \ref{Klein_Gordon_Act}. By introducing new bosonic operators
\begin{align}
a_{\bm{k}}=&\frac{1}{2}X_{\bm{k}}-\frac{1}{\sqrt{2\epsilon_{\bm{k}}}}\Pi^\dagger_{\bm{k}},\quad a^\dagger_{\bm{k}}=\frac{1}{2}X^\dagger_{\bm{k}}+\frac{1}{\sqrt{2\epsilon_{\bm{k}}}}\Pi_{\bm{k}},\\
b_{\bm{k}}=&\frac{1}{2}X^\dagger_{\bm{k}}-\frac{1}{\sqrt{2\epsilon_{\bm{k}}}}\Pi_{\bm{k}},\quad b^\dagger_{\bm{k}}=\frac{1}{2}X_{\bm{k}}+\frac{1}{\sqrt{2\epsilon_{\bm{k}}}}\Pi^\dagger_{\bm{k}},
\end{align}
we transform the relativistic Klein Gordon action to the nonrelativistic Schrodinger type
\begin{align}\nonumber
S_\phi=&\int_0^\beta\sum_{\bm{k}, \bm{q}}\left[a_{\bm{k}}\left(-\partial_\tau\delta_{\bm{k}, \bm{q}}+iA_{0, -\bm{k}+\bm{q}}\right)a^\dagger_{\bm{q}}+b^\dagger_{\bm{k}}\left(\partial_\tau\delta_{\bm{k}, \bm{q}}\right.\right.\\
&\left.\left.-iA_{0, -\bm{k}+\bm{q}}\right)b_{\bm{q}}+\epsilon_{\bm{k}}\left(a_{\bm{k}}a^\dagger_{\bm{k}}+b^\dagger_{\bm{k}}b_{\bm{k}}\right)\right].
\end{align}
Importantly, here $a^{\left(\dagger\right)}_{\bm{k}}$ is the holon operator and $b^{\left(\dagger\right)}_{\bm{k}}$ is the doublon operator. The doublon and holon together composes the relativistic chargon $X_{\bm{r}}=a_{\bm{r}}+b^\dagger_{\bm{r}}$, $X^\dagger_{\bm{r}}=a^\dagger_{\bm{r}}+b_{\bm{r}}$.

Now with only longitudinal gauge field, the $U\left(1\right)$ QSL with SFS action $S_{\textrm{SL}}$ takes the form
\begin{align}\nonumber
S_{\textrm{SL}}=&\int_0^\beta d\tau\sum_{\sigma, \bm{k}, \bm{q}}\left[f^\dagger_{\sigma, \bm{k}}\left(\partial_\tau+\xi_{\bm{k}}\right)f_{\sigma, \bm{k}}+a_{\bm{k}}\left(-\partial_\tau+\epsilon_{\bm{k}}\right)a^\dagger_{\bm{k}}\right.\\\nonumber
&\left.+b^\dagger_{\bm{k}}\left(\partial_\tau+\epsilon_{\bm{k}}\right)b_{\bm{k}}+if^\dagger_{\sigma, \bm{k}}A_{0, \bm{k}-\bm{q}}f_{\sigma, \bm{q}}+ia_{\bm{k}}A_{0, -\bm{k}+\bm{q}}a^\dagger_{\bm{q}}\right.\\
&\left.-ib^\dagger_{\bm{k}}A_{0, -\bm{k}+\bm{q}}b_{\bm{q}}+\frac{\bm{q}^2}{2g^2\Omega}A_{0, -\bm{q}}A_{0, \bm{q}}\right].
\end{align}
Integrating out the longitudinal gauge field $\bm{A}_{0, \bm{r}}$ then yields the gauge field $A_{0, \bm{r}}$ fluctuations induced interaction terms
\begin{align}\nonumber
&\frac{1}{2}\int d\bm{r}\int d\bm{r}'U^{\textrm{C}}_{\bm{r}-\bm{r}'}\sum_{\sigma, \sigma'}\left[f^\dagger_{\sigma, \bm{r}}f_{\sigma, \bm{r}}f^\dagger_{\sigma, ', \bm{r}'}f_{\sigma', \bm{r}'}+\left(a^\dagger_{\bm{r}}a_{\bm{r}}\right.\right.\\
-&\left.\left.b^\dagger_{\bm{r}}b_{\bm{r}}\right)\left(a^\dagger_{\bm{r}'}a_{\bm{r}'}-b^\dagger_{\bm{r}'}b_{\bm{r}'}\right)+2f^\dagger_{\sigma, \bm{r}}f_{\sigma, \bm{r}}\left(a^\dagger_{\bm{r}'}a_{\bm{r}'}-b^\dagger_{\bm{r}'}b_{\bm{r}'}\right)\right],
\end{align}
where the Coulomb type interaction $U^{\textrm{C}}_{\bm{r}-\bm{r}'}=\frac{g^2}{\left(2\pi\right)^2}\int\frac{e^{i\bm{q}\cdot\left(\bm{r}-\bm{r}'\right)}}{\bm{q}^2+K^2}d\bm{q}$ has the screening $K$ due to the itinerant spinons~\cite{Cody}. The interaction arising from longitudinal gauge field fluctuations includes the spinon spinon interaction, the chargon chargon interaction, and the spinon chargon binding interaction. The spinon spinon interaction and the chargon chargon interaction modify the spinon and chargon correlation function respectively, but the last term, the spinon chargon binding interaction tends to bind the spinon and the chargon into an electron, which becomes the dominant effect of the gauge field fluctuations~\cite{Tang}. 

Taking into the dominant effect of the gauge field fluctuations into account, we arrive at the QSL action in the presence of longitudinal gauge field fluctuations
\begin{align}\nonumber
S_{\textrm{SL}}\approx&\int_0^\beta d\tau\sum_{\sigma, \bm{k}}\left[f^\dagger_{\sigma, \bm{k}}\left(\partial_\tau+\xi_{\bm{k}}\right)f_{\sigma, \bm{k}}+a_{\bm{k}}\left(-\partial_\tau+\epsilon_{\bm{k}}\right)a^\dagger_{\bm{k}}\right.\\
&\left.+b^\dagger_{\bm{k}}\left(\partial_\tau+\epsilon_{\bm{k}}\right)b_{\bm{k}}+\int U_{\bm{r}}f^\dagger_{\sigma, \bm{r}}f_{\sigma, \bm{r}}\left(a^\dagger_{\bm{r}}a_{\bm{r}}-b^\dagger_{\bm{r}}b_{\bm{r}}\right)d\bm{r}\right].
\end{align}
Here, we only consider the onsite spinon chargon binding interaction since the the screening from the itinerant spinons makes the interaction short range. We know that the Thomas Fermi screening is $K=g\sqrt{N_f}$ with $N_f$ being the spinon DOS at the Fermi energy, so at $\bm{q}\rightarrow\bm{0}$ the spinon chargon binding interaction strength is approximated as
\begin{align}
U_{\bm{r}}\approx&\frac{g^2}{\left(2\pi\right)^2}\int\frac{1}{g^2N_f}d\bm{q}=\int\frac{d\bm{q}}{4\pi^2N_f}\approx\Lambda_f.
\end{align}
Here $\Lambda_f$ is half of the spinon band width. It indicates that the the spinon chargon binding interaction $U_{\bm{r}}$ is estimated to be half of the spinon band width.

\section{DERIVATION OF GAUGE FIELDS FLUCTUATIONS MODIFIED IMPURITY ELECTRONIC GREEN'S FUNCTION}\label{App-F}
By numerically solving the self-consistent equations in Eq. \ref{Num_SF_Eq1} - \ref{Num_SF_Eq4}, we can obtain the SAF spionon coupling $w$, the CB chargon coupling $u$, and the Lagrangian multipliers $\lambda$, $h$, so the magnetic impurity on the $U\left(1\right)$ QSL with SFS has the quadratic action at the mean field level
\begin{widetext}
\begin{align}\nonumber
S_0=&\int_0^\beta\sum_\sigma\left[a^\dagger_\sigma\left(\partial_\tau+\epsilon_0-h\right)a_\sigma+w\left(f^\dagger_{\sigma, \bm{R}}a_\sigma+a^\dagger_\sigma f_{\sigma, \bm{R}}\right)\right]d\tau\\\nonumber
&+\int_0^\beta\left[\frac{1}{U}\left(\partial_\tau+h\right)X^\dagger_d\left(\partial_\tau-h\right)X_d+\left(\lambda+\frac{h^2}{U}\right)X^\dagger_dX_d-u\left(X^\dagger_d X_{\bm{R}}+X^\dagger_{\bm{R}}X_d\right)\right]d\tau\\
&-\int_0^\beta d\tau\int_0^\beta d\tau'\sum_\sigma\left[f^\dagger_{\sigma, \bm{R}}\left(\tau\right)G^{-1}_{f, \sigma}\left(\tau, \tau', \bm{R}, \bm{R}\right)f_{\sigma, \bm{R}}\left(\tau'\right)-X^\dagger_{\bm{R}}\left(\tau\right)G^{-1}_{X}\left(\tau, \tau', \bm{R}, \bm{R}\right)X_{\bm{R}}\left(\tau'\right)\right].
\end{align}
\end{widetext}
In order to incorporate the spinon chargon binding interaction in the QSL into the mean field action $S_0$, the holon and doublon operators at the impurity are defined as
\begin{align}
a_d=&\frac{1}{2}\left(X_d-\frac{U}{\sqrt{\lambda U+h^2}}\Pi^\dagger_d\right),\\
a^\dagger_d=&\frac{1}{2}\left(X^\dagger_d+\frac{U}{\sqrt{\lambda U+h^2}}\Pi_d\right),\\
b_d=&\frac{1}{2}\left(X^\dagger_d-\frac{U}{\sqrt{\lambda U+h^2}}\Pi_d\right),\\
b^\dagger_d=&\frac{1}{2}\left(X_d+\frac{U}{\sqrt{\lambda U+h^2}}\Pi^\dagger_d\right),
\end{align}
with the canonical momentum operators defined to be
\begin{align}
\Pi^\dagger_d=\frac{\left(\partial_\tau-h\right)X_d}{U},\quad \Pi_d=\frac{\left(\partial_\tau+h\right)X^\dagger_d}{U}.
\end{align}
In this way, the CB at the impurity, which is also relativistic, is expressed by the impurity holon and doublon operators as $X_d=a_d+b^\dagger_d$, $X^\dagger_d=a^\dagger_d+b_d$. Therefore, the relativistic type Klein Gordon field action at the impurity can also be transformed to the Schrodinger field type. 

In the Matsubara frequency space, the Green's functions for the holon and doublon in the QSL are
\begin{align}
G_a\left(i\nu_n, \bm{R}, \bm{R}\right)=&\sum_{\bm{k}}\frac{1}{i\nu_n-\epsilon_{\bm{k}}},\\
G_{b}\left(i\nu_n, \bm{R}, \bm{R}\right)=&\sum_{\bm{k}}\frac{-1}{i\nu_n+\epsilon_{\bm{k}}},
\end{align}
and the Matsubara Green's functions for the holon and doublon at the impurity take the form
\begin{align}
G_{a_d}\left(i\nu_n\right)=&\sqrt{\frac{U}{2\sqrt{\lambda U+h^2}}}\frac{1}{i\nu_n-\sqrt{\lambda+\frac{h^2}{U}}+h},\\
G_{b_d}\left(i\nu_n\right)=&\sqrt{\frac{U}{2\sqrt{\lambda U+h^2}}}\frac{-1}{i\nu_n+\sqrt{\lambda+\frac{h^2}{U}}+h}.
\end{align}
With the holon and doublon operators in both the impurity and the QSL, the mean field action $S_0$ now becomes
\begin{widetext}
\begin{align}\nonumber\label{tilde_act}
S_0=&-\sum_{\sigma, \omega_n}\begin{pmatrix}
a^\dagger_\sigma\left(i\omega_n\right) & f^\dagger_{\sigma, \bm{R}}\left(i\omega_n\right)
\end{pmatrix}\begin{pmatrix}
i\omega_n-\epsilon_0+h & -w \\
-w & G^{-1}_{f, \sigma}\left(i\omega_n, \bm{R}, \bm{R}\right)
\end{pmatrix}
\begin{pmatrix}
a_\sigma\left(i\omega_n\right) \\ f_{\sigma, \bm{R}}\left(i\omega_n\right)
\end{pmatrix}\\
&-\sum_{\nu_n}\begin{pmatrix}
a_d\left(i\nu_n\right) & b^\dagger_d\left(i\nu_n\right) & a_{\bm{R}}\left(i\nu_n\right) & b^\dagger_{\bm{R}}\left(i\nu_n\right)
\end{pmatrix}\begin{pmatrix}
G^{-1}_{a_d}\left(i\nu_n\right) & 0 & -u & -u \\
0 & G^{-1}_{b_d}\left(i\nu_n\right) & -u & -u \\
-u & -u & G^{-1}_a\left(i\nu_n\right) & 0 \\
-u & -u & 0 & G^{-1}_b\left(i\nu_n\right)
\end{pmatrix}\begin{pmatrix}
a^\dagger_d\left(i\nu_n\right) \\ b_d\left(i\nu_n\right) \\ a^\dagger_{\bm{R}}\left(i\nu_n\right) \\ b_{\bm{R}}\left(i\nu_n\right)
\end{pmatrix}.
\end{align}
\end{widetext}

\begin{figure}
\centering
\includegraphics[width=3.5in]{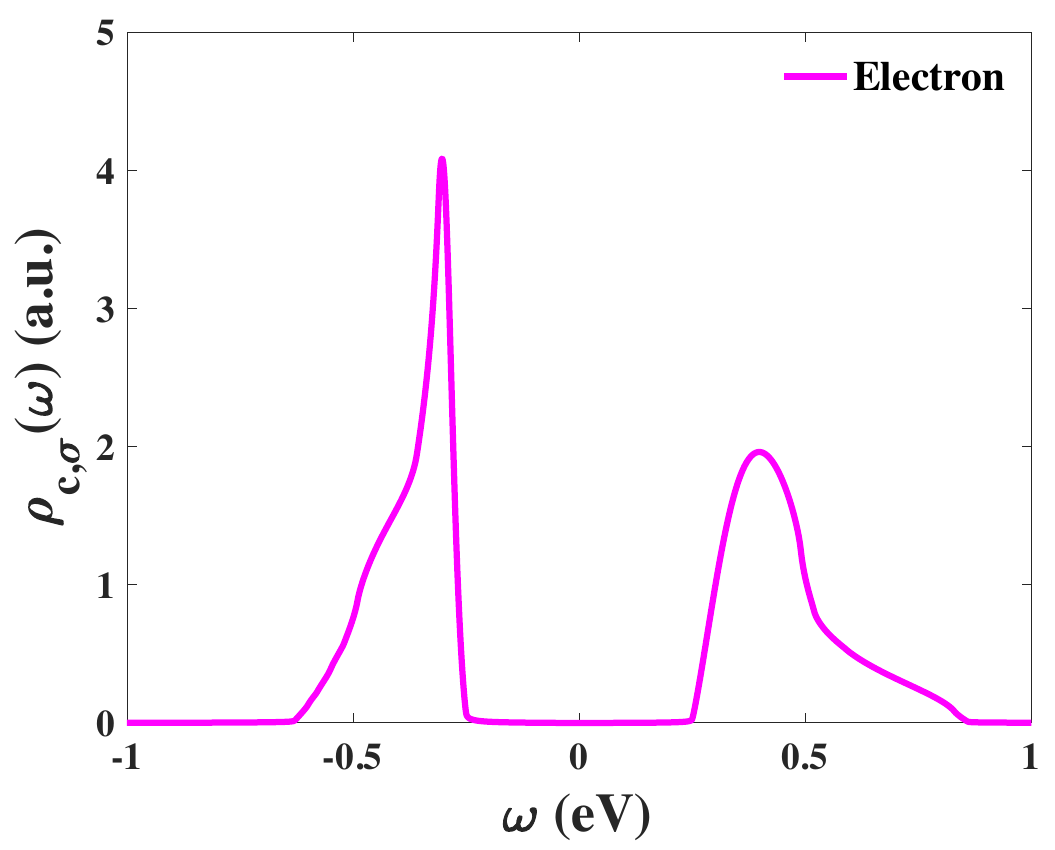}
\caption{The electronic spectral function $\rho_{c, \sigma}\left(\omega\right)$ in the QSL. The gauge field fluctuations induced spinon chargon binding interaction is taken to be $U_{\bm{R}}=0.23$ eV.}\label{figureS2}
\end{figure}

In the slave rotor representation, the electron at the impurity is composed of the SAF combined with the impurity holon and doublon
\begin{align}
d^\dagger_\sigma=a^\dagger_\sigma\left(a_d+b^\dagger_d\right),\quad d_\sigma=a_\sigma\left(a^\dagger_d+b_d\right).
\end{align}
The electrons in the QSL are composed of the spinons combined with the QSL holons and doublons
\begin{align}
c^\dagger_{\sigma, \bm{r}}=f^\dagger_{\sigma, \bm{r}}\left(a_{\bm{r}}+b^\dagger_{\bm{r}}\right),\quad c_{\sigma, \bm{r}}=f_{\sigma, \bm{r}}\left(a^\dagger_{\bm{r}}+b_{\bm{r}}\right).
\end{align}
With the holon and doublon operators, the electronic bare Matsubara Green's function that involves the impurity, the nearest neighboring QSL site, and the coupling is constructed as
\begin{align}
G^0_\sigma\left(i\omega_n\right)=&\begin{pmatrix}
\hat{G}^0_{d, \sigma}\left(i\omega_n\right) & \hat{T}_\sigma\left(i\omega_n\right) \\
\hat{T}^\dagger_\sigma\left(i\omega_n\right) & \hat{G}_{c, \sigma}\left(i\omega_n\right)
\end{pmatrix},
\end{align}
with
\begin{widetext}
\begin{align}
\hat{G}^0_{d, \sigma}\left(i\omega_n\right)=&-\frac{1}{\beta}\sum_{\nu_n}\left\langle a_\sigma\left(i\omega_n+i\nu_n\right)a^\dagger_\sigma\left(i\omega_n+i\nu_n\right) \right\rangle_0\left\langle \begin{pmatrix}
a^\dagger_d\left(i\nu_n\right)a_d\left(i\nu_n\right) & a^\dagger_d\left(i\nu_n\right)b^\dagger_d\left(i\nu_n\right) \\
b_d\left(i\nu_n\right)a_d\left(i\nu_n\right) & b_d\left(i\nu_n\right)b^\dagger_d\left(i\nu_n\right)
\end{pmatrix}
\right\rangle_0,
\end{align}
\begin{align}
\hat{T}_\sigma\left(i\omega_n\right)=&-\frac{1}{\beta}\sum_{\nu_n}\left\langle a_\sigma\left(i\omega_n+i\nu_n\right)f^\dagger_{\sigma, \bm{R}}\left(i\omega_n+i\nu_n\right) \right\rangle_0\left\langle \begin{pmatrix}
a^\dagger_d\left(i\nu_n\right)a_{\bm{R}}\left(i\nu_n\right) & a^\dagger_d\left(i\nu_n\right) b^\dagger_{\bm{R}}\left(i\nu_n\right) \\
b_d\left(i\nu_n\right) a_{\bm{R}}\left(i\nu_n\right) & b_d\left(i\nu_n\right)b^\dagger_{\bm{R}}\left(i\nu_n\right)
\end{pmatrix} \right\rangle_0,
\end{align}
\begin{align}
\hat{T}^\dagger_\sigma\left(i\omega_n\right)=&-\frac{1}{\beta}\sum_{\nu_n}\left\langle f_{\sigma, \bm{R}}\left(i\omega_n+i\nu_n\right)a^\dagger_\sigma\left(i\omega_n+i\nu_n\right) \right\rangle_0\left\langle\begin{pmatrix}
a^\dagger_{\bm{R}}\left(i\nu_n\right)a_d\left(i\nu_n\right) & a^\dagger_{\bm{R}}\left(i\nu_n\right)b^\dagger_d\left(i\nu_n\right) \\
b_{\bm{R}}\left(i\nu_n\right)a_d\left(i\nu_n\right) & b_{\bm{R}}\left(i\nu_n\right)b^\dagger_d\left(i\nu_n\right)
\end{pmatrix}\right\rangle_0,
\end{align}
\begin{align}
\hat{G}_{c, \sigma}\left(i\omega_n\right)=&-\frac{1}{\beta}\sum_{\nu_n}\left\langle f_{\sigma, \bm{R}}\left(i\omega_n+i\nu_n\right)f^\dagger_{\sigma, \bm{R}}\left(i\omega_n+i\nu_n\right)\right\rangle_0\left\langle\begin{pmatrix} 
a^\dagger_{\bm{R}}\left(i\nu_n\right)a_{\bm{R}}\left(i\nu_n\right) & a^\dagger_{\bm{R}}\left(i\nu_n\right)b^\dagger_{\bm{R}}\left(i\nu_n\right) \\
b_{\bm{R}}\left(i\nu_n\right)a_{\bm{R}}\left(i\nu_n\right) & b_{\bm{R}}\left(i\nu_n\right)b^\dagger_{\bm{R}}\left(i\nu_n\right)
\end{pmatrix}\right\rangle_0.
\end{align}
\end{widetext}
Those $2\times2$ correlators can all be obtained from the action $S_0$ in Eq. \ref{tilde_act}. In this way the impurity bare electronic Matsubara Green's function $G^0_{d, \sigma}\left(i\omega_n\right)$ is recovered as $G^0_{d, \sigma}\left(i\omega_n\right)=\sum_{i, j=1, 2}\left[\hat{G}^0_{d, \sigma}\left(i\omega_n\right)\right]_{ij}$.

The spinon chargon binding interaction in the QSL introduces new terms to the action
\begin{align}
S'=\int_0^\beta\int U_{\bm{r}}\sum_\sigma\left(f^\dagger_{\sigma, \bm{r}}f_{\sigma, \bm{r}}a^\dagger_{\bm{r}}a_{\bm{r}}-f^\dagger_{\sigma, \bm{r}}f_{\sigma, \bm{r}}b^\dagger_{\bm{r}}b_{\bm{r}}\right)d\bm{r} d\tau,
\end{align}
and now the total action is $S=S_0+S'$. Taking the effect of the spinon chargon binding interaction into account, we can use the random phase approximation to calculate the Matsubara electronic Green's function. The resulting Matsubara electronic Green's function is
\begin{align}
G_\sigma\left(i\omega_n\right)=\left[1-G_{0, \sigma}\left(i\omega_n\right)\hat{U}\right]^{-1}G_{0, \sigma}\left(i\omega_n\right),
\end{align}
with
\begin{align}
\hat{U}=\begin{pmatrix}
0 & 0 & 0 & 0 \\
0 & 0 & 0 & 0 \\
0 & 0 & U_{\bm{R}} & 0 \\
0 & 0 & 0 & -U_{\bm{R}}
\end{pmatrix}.
\end{align}
As a result, the spinon chargon binding interaction modified impurity electronic Green's function can be eventually obtained as
\begin{widetext}
\begin{align}
G_{d, \sigma}\left(i\omega_n\right)=&\sum_{i, j=1, 2}\left[G_\sigma\left(i\omega_n\right)\right]_{ij}=G^0_{d, \sigma}\left(i\omega_n\right)+\sum_{i, j=1, 2}\left\{\hat{T}_\sigma\left(i\omega_n\right)U_{\bm{R}}\sigma_z\left[1-\hat{G}_{c, \sigma}\left(i\omega_n\right)U_{\bm{R}}\sigma_z\right]^{-1}\hat{T}_\sigma\left(i\omega_n\right)\right\}_{ij},
\end{align}
and the QSL electronic Green's function is
\begin{align}
G_{c, \sigma}\left(i\omega_n\right)=\sum_{i, j=1, 2}\left\{\hat{G}_{c, \sigma}\left(i\omega_n\right)\left[1-\hat{G}_{c, \sigma}\left(i\omega_n\right)U_{\bm{R}}\sigma_z\right]^{-1}\right\}_{ij}.
\end{align}
\end{widetext}
Given the same onsite spinon chargon binding interaction $U_{\bm{R}}=0.23$ eV, the QSL electronic spectral function $\rho_{c, \sigma}\left(\omega\right)=-\frac{1}{\pi}\textrm{Im}G_{c, \sigma}\left(i\omega_n\rightarrow\omega+i0^+\right)$ is plotted in Fig. \ref{figureS2}. It can be seen in Fig. \ref{figureS2} that the onsite spinon chargon binding interaction changes the QSL electronic spectral function $\rho_{c, \sigma}\left(\omega\right)$ compared with that in Fig. \ref{figureS1} (c), but there is no resonance peaks emerging at the Hubbard band edges.


\bibliographystyle{apsrev4-1} 

\end{document}